\documentclass[10pt, doublecolumn]{IEEEtran}
\usepackage{booktabs}
\usepackage[caption=false, font=footnotesize]{subfig}
\newcommand*{\rom}[1]{\expandafter\@slowromancap\romannumeral #1@}
\usepackage{amsmath, amssymb}
\usepackage{xcolor}
\usepackage{cite}
\usepackage{enumitem}
\usepackage{url}
\usepackage{bbm}

\ifCLASSINFOpdf
   \usepackage[pdftex]{graphicx}
   \graphicspath{{Figures/}}
\else
\fi
\newcommand{\PMC}{p_{m}}
\newcommand{\SA}{s_{m}}

\begin{document}
\bstctlcite{IEEEexample:BSTcontrol}
\font\myfont=cmr16 at 23pt

\title{Balancing AoI and Rate for Mission-Critical and eMBB Coexistence with Puncturing, NOMA,
and RSMA in Cellular Uplink}

\author{
        Farnaz~Khodakhah,~\IEEEmembership{Student~Member,~IEEE,}
        Aamir~Mahmood,~\IEEEmembership{Senior~Member,~IEEE,}\\
        \v{C}edomir~Stefanovi\'{c},~\IEEEmembership{Senior~Member,~IEEE,}
        Hossam~Farag,~\IEEEmembership{Member,~IEEE,}
        Patrik~{\"O}sterberg,~\IEEEmembership{Senior~Member,~IEEE,}
       and Mikael~Gidlund,~\IEEEmembership{Senior~Member,~IEEE}%
\thanks{F.~Khodakhah, A.~Mahmood, P.~{\"O}sterberg and M.~Gidlund are with the Department of Computer and Electrical Engineering, Mid Sweden University, 851 70 Sundsvall, Sweden  (e-mail: \{farnaz.khodakhah, aamir.mahmood, patrik.osterberg, mikael.gidlund\}@miun.se).
}%
\thanks{\v{C}.~Stefanovi\'{c}, H.~Farag are with the Department of Electronic Systems, Aalborg University, 2450 K\o benhavn SV, Denmark (e-mail: \{cs,hmf\}@es.aau.dk).}%
\vspace{-22pt}
}%
\maketitle
\thispagestyle{empty}
\begin{abstract}
Through the lens of average and peak age-of-information (AoI), this paper takes a fresh look into the uplink medium access solutions for mission-critical (MC) communication coexisting with enhanced mobile broadband (eMBB) service. Considering the stochastic packet arrivals from an MC user, we study three access schemes: 
orthogonal multiple access (OMA) with eMBB preemption (puncturing), non-orthogonal multiple access (NOMA), and rate-splitting multiple access (RSMA), the latter two both with concurrent eMBB transmissions.
Puncturing is found to reduce both average AoI and peak AoI (PAoI) violation probability but at the expense of decreased eMBB user rates and increased signaling complexity.
Conversely, NOMA and RSMA offer higher eMBB rates but may lead to MC packet loss and AoI degradation.
The paper systematically investigates the conditions under which NOMA or RSMA can closely match the average AoI and PAoI violation performance of puncturing while maintaining data rate gains.
Closed-form expressions for average AoI and PAoI violation probability are derived, and conditions on the eMBB and MC channel gain difference with respect to the base station are analyzed.
Additionally, optimal power and rate splitting factors in RSMA are determined through an exhaustive search to minimize MC outage probability.
Notably, our results indicate that with a small loss in the average AoI and PAoI violation probability the eMBB rate in NOMA and RSMA can be approximately five times higher than that achieved through puncturing.

\end{abstract}

\begin{IEEEkeywords}
NOMA, RSMA, puncturing, AoI, heterogeneous services, eMBB, URLLC, MC, PAoI.
\end{IEEEkeywords}

\section{Introduction}
\label{Intro}
The services offered by 5G technology are categorized into three primary groups: enhanced mobile broadband (eMBB), ultra reliable low latency (URLLC), and massive machine-type communication (mMTC).
eMBB services demand reliable connections characterized by exceptionally high peak data rates, along with moderate rates catering to users at the cell edges.
URLLC services prioritize low-latency and age-sensitive transmissions coupled with high reliability.
Finally, mMTC services are designed to support a vast number of users that exhibit sporadic activity.

In light of the diverse service requirements, particularly in the emerging application scenarios with the likely coexistence of eMBB, URLLC, and mMTC services, there is a growing need to reevaluate/redesign the radio access network (RAN) architecture and access solutions in 5G-and-beyond systems~\cite{i2016new}.
To address the challenge of providing customized, reliable services within the constraints of limited network resources, while also minimizing capital and operating expenses, 
network slicing is envisioned as a primary enabler.
Network slicing involves segmenting a physical network into multiple logical networks, enabling the support of on-demand services tailored to distinct application scenarios, all within the framework of the same physical network\cite{rost2016mobile,jiang2016network,zhang2017network}.
The key idea is to perform dynamic and efficient allocation of network resources to such created slices, aligning with specific quality of service (QoS) demands. At RAN, slicing for multi-service coexistence scenarios can be achieved by
\begin{itemize}[leftmargin=*]
    \item \textit{Orthogonal Multiple Access (OMA) based slicing}, which allocates mutually isolated time-frequency resources of the radio link to the users. 
    \item \textit{Non Orthogonal Multiple Access (NOMA) based slicing}, wherein the users exploit the same time-frequency resources while maintaining separation in power and/or spatial domains. 
    \item \textit{Rate Splitting Multiple Access (RSMA) based slicing}, like NOMA-based slicing, allows users to share the same time-frequency resources by, for instance, power-domain separation.  However, in RSMA-based non-orthogonal slicing, the users can split their individual messages into two streams and apply different power- and rate-splitting factors to each before transmitting a superposed message concurrently with other users in the system~\cite{mishra2022rate,rimoldi1996rate}.
    By doing so, RSMA can be used both uplink and downlink~\cite{mao2022rate,mishra2022rate} and is shown to be sum-rate optimal in certain conditions compared to NOMA~\cite{mao2018energy}, thus gaining the status of the next-generation multiple access solution. 
\end{itemize}

In this paper, we focus on analyzing the coexistence performance of eMBB and mission-critical (MC) services, with a specific emphasis on identifying a well-balanced RAN slicing (in terms of medium access) solution, given the necessity to support mixed-traffic scenarios.
Note that MC is closely related to the URLLC service category, requiring a timely and reliable exchange of information flows within the serviced applications. 
MC applications can be found across various industries, such as industrial automation, remote robot control, and autonomous vehicle coordination~\cite{mahmood2021industrial}, where ensuring the freshness of status updates is paramount; otherwise, relying on outdated information can lead to undesired state-action performance~\cite{hydher2020intelligent,liu2018age,abd2018average}.
In this context, the age-of-information (AoI) serves as a suitable metric that combines timeliness and reliability and quantifies the freshness of received information.
AoI captures the time elapsed since the last successfully received piece of information, typically the application layer message~\cite{liu2018age}.
Previous studies have explored the suitability of AoI as a metric for MC applications~\cite{basnayaka2021age,sun2017update,yavascan2021analysis}. 
Meanwhile, Costa \textit{et al.}~\cite{costa2016age} and He \textit{et al.}~\cite{he2016optimal} introduced a derivative metric, the peak age-of-information (PAoI), which characterizes the maximum value of the age immediately before receiving a new packet.
PAoI offers closer insights into the worst-case performance and can be used to formulate performance guarantees in a probabilistic sense, as opposed to relying on average AoI.

For eMBB-MC coexistence, the focus of this paper is on an uplink scenario in which the eMBB service is characterized by the full buffer model, while the MC service is characterized by sporadic and brief transmissions.
We investigate OMA-, NOMA-, and RSMA-based slicing setups.\footnote{Precisely speaking, we consider heterogeneous OMA (H-OMA) and heterogeneous NOMA (H-NOMA) in this work, as the multiplexed traffic is of heterogeneous nature.}
In regards to the former, we assume that OMA-based slicing takes the form of preempting (puncturing), such that the base station (BS) temporarily punctures eMBB transmission during short MC transmissions. 
On the other hand, in NOMA- and RSMA-based schemes, when activated the MC user transmits concurrently with the eMBB user.
\textit{Our goal is to identify the conditions under which a strategic choice among the schemes can be made, such that the freshness of the delivered information by the MC user can be kept close to the best possible one, while simultaneously maximizing the rate for the eMBB user.}

The paper extends our preliminary work~\cite{khodakhah2023noma}, where we analyzed and compared only the puncturing and NOMA schemes.
The findings from \cite{khodakhah2023noma} indicate that puncturing offers a superior average AoI performance for the MC service at the cost of reducing the eMBB service data rate.
We also identified the conditions under which the use of NOMA incurs only a marginal penalty in the average AoI performance for MC service while significantly increasing the achievable rate for the eMBB user.
A limitation of this preliminary work is that we focused on scenarios with fixed eMBB and MC locations, and the switching strategy depends on the distance of the users from the base station (BS).
In this paper, we base our analysis on the signal-to-noise ratio (SNR) gap between the eMBB and MC users; this metric reflects the channel gain difference between the users, allowing us to derive generic conditions for the transition from one medium access strategy to another.
In addition, besides the average AoI performance, in this work, we also include \emph{PAoI violation probability} (i.e., the probability that PAoI will exceed a certain limit) for the MC user in our investigations. In summary, our key contributions in this work are:

\begin{itemize}
    \item To the best of our knowledge, this paper is the first to investigate how the AoI-based metrics perform for different access solutions, especially in the context of the RSMA scheme. To compare with other solutions, we determine the optimal power and rate splitting factors for the RSMA scheme, such that the performance metrics are optimized. We provide important insights into how the choice of the access scheme influences the dynamics and effectiveness of information freshness.
    \item We develop a discrete-time Markov chain (DTMC) model to evaluate both the average AoI performance and the PAoI violation probability for the MC user, 
    encompassing the puncturing, NOMA, and RSMA approaches.
    \item Given a tolerance for slight gaps in the average AoI performance and in the PAoI violation probability in comparison to the puncturing approach, we determine the conditions favoring the use of NOMA/RSMA such that the eMBB user rate is maximized. We investigate how these conditions change depending on the MC user activation probability.
\end{itemize}

%
The rest of the paper is organized as follows. 
Sec.~\ref{Literature} positions this paper with respect to the literature.
Sec.~\ref{System.Model} introduces the system model, and 
Sec.~\ref{Analysis} derives the key metrics of interest. 
Optimization of the choice of the access schemes is presented in Sec.~\ref{Sec.Optimization}.
Sec.~\ref{Evaluation} is dedicated to performance results and analysis.
Finally, Sec.~\ref{Conclusion} summarizes our main findings.

\section{Related Work}
\label{Literature}
When exploring related works, our primary focus centers on scenarios involving uplink multi-class traffic coexistence. However, for comprehensive coverage, we also acknowledge relevant downlink studies. Given our objective of analyzing concurrent transmissions of eMBB and MC users within a time-frequency resource pool, we consider works related to AoI and PAoI, despite their emphasis on eMBB-URLLC coexistence on orthogonal time-frequency resources, which falls beyond the scope of this section.

NOMA-based slicing schemes in uplink scenarios have been extensively studied, as evidenced by the works of Popovski et al. \cite{popovski20185g} and Tominaga et al. \cite{tominaga2021non}. Popovski et al. \cite{popovski20185g} investigated the coexistence of eMBB and URLLC, demonstrating that Heterogeneous NOMA (H-NOMA) outperforms Heterogeneous Orthogonal Multiple Access (H-OMA) under certain conditions. Specifically, when the URLLC rate exceeds that of eMBB, H-NOMA showcases superior performance due to the BS's ability to decode and eliminate URLLC transmissions through reliability diversity. Conversely, in scenarios where the URLLC rate is lower than that of eMBB, H-OMA proves advantageous, achieving eMBB and URLLC pairs unattainable by H-NOMA with successive interference cancellation (SIC). This preference for H-OMA arises from the challenge of maintaining high reliability during eMBB transmissions, especially when the URLLC channel gain is lower than that of eMBB.

Conversely, if the goal is to ensure high eMBB sum rates, H-NOMA provides significant performance improvements. Non-orthogonal transmission enables eMBB users to utilize a greater number of spectral resources without substantial interference from URLLC. Similarly, Tominaga et al. \cite{tominaga2021non} demonstrated the superiority of NOMA-based slicing over OMA-slicing under certain channel condition scenarios.

Despite these advancements, the assessment of conditions where H-NOMA outperforms H-OMA while meeting reliability and latency constraints for eMBB users remains relatively unexplored in existing literature. Additionally, studies like that of \cite{CLSKP2022} explore similar scenarios, investigating the coexistence of broadband and intermittently active users in the cellular uplink with various access methods.

Furthermore, RSMA has garnered significant attention for both downlink and uplink communications \cite{mao2022rate,mishra2022rate}. In the downlink, RSMA enables the transmission of common and private symbols decoded by multiple users, offering a robust framework for non-orthogonal transmission and interference management \cite{clerckx2019rate}. In the uplink, RSMA involves splitting a user's message into two streams and transmitting a superposed message to the BS, where power allocation between streams critically impacts performance \cite{mao2022rate,rimoldi1996rate}.

By optimizing decoding order and power allocation, RSMA achieves the capacity region between boundary points with SIC without time-sharing \cite{rimoldi1996rate,yang2020sum}. In addition, RSMA has shown promise in enhancing fairness among users, improving outage performance, and simplifying uplink implementation by eliminating the need for user pairing \cite{liu2020rate,liu2021rate,zhu2017rate}. Studies such as \cite{dos2021rate} have illustrated RSMA's potential to achieve larger rate regions than NOMA and OMA-based slicing under certain conditions.

However, the exploration of RSMA in scenarios where neither NOMA nor OMA serve as suitable alternatives remains largely unaddressed. This gap arises from the complexity RSMA introduces to the system in terms of decoding and transmission processes. Furthermore, no prior studies have extensively explored uplink RSMA in the context of AoI in the coexistence of heterogeneous services. Our paper aims to fill this gap by evaluating uplink RSMA's performance for average and peak AoI of MC users, showcasing its feasibility and effectiveness in meeting MC application requirements.

\section{System Model}
\label{System.Model}

\begin{figure*}
    \centering
    \includegraphics[width=1\linewidth]{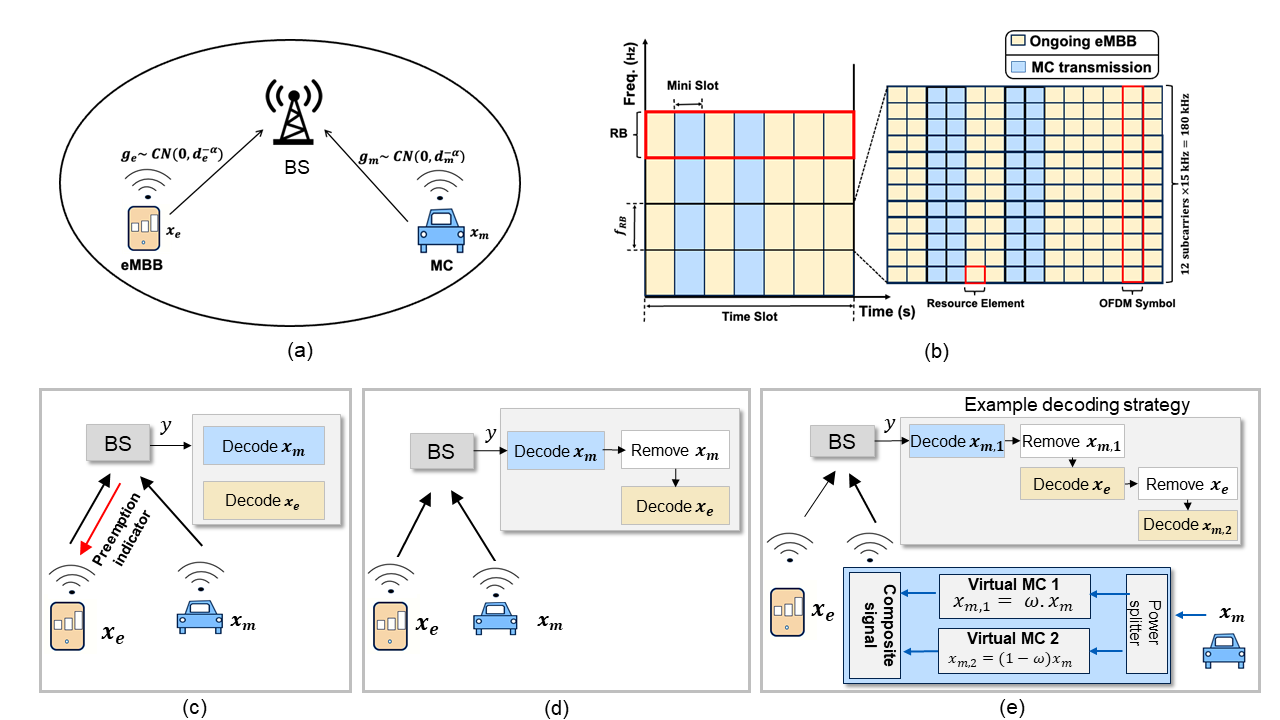}
    \caption{System model of eMBB-MC coexistence: (a) Single-cell uplink transmission scenario of eMBB and MC users, (b) Time-frequency resource grid with prescheduled eMBB user and generic overlapping access scheme for a ``generate-at-will" MC user, (c) Puncturing-based access, (d) NOMA-based access, and (e) RSMA-based access scheme.}
    \label{fig:enter-label}
\end{figure*}

We consider an uplink 5G communication scenario with one eMBB and one MC user, denoted by $e$ and $m$ respectively and located at distance $d_e$ and $d_m$ from a common BS, as shown in Fig.~\ref{fig:enter-label}(a).
The time-frequency grid is divided into $B$ resource blocks (RBs), 
where a RB occupies 12 sub-carriers~\cite{3GPP.TS.38.211}, 
with the subcarrier spacing of 15\,kHz and the system bandwidth of $BW = f_\text{RB} \, B$, where $f_\text{RB}$ is the bandwidth of a RB, as shown in Fig.~\ref{fig:enter-label}(b). 
The link-time is divided into time slots of $1$\,ms duration, each time slot contains $\mathcal{S} = \{1,2,...,S\}$, where $S$ represents the total number of minislots, with each minislot containing $n_\text{sym}$ OFDM symbols.
We consider the full-buffer traffic model for the eMBB user. The full-buffer traffic model for the eMBB user refers to the scenario where there is always a packet in the queue waiting to be transmitted. It implies that there is always an eMBB packet scheduled in the time-frequency resource grid, experiencing coexisting MC user traffic. This model is commonly employed by 3GPP (3rd Generation Partnership Project) to simulate high traffic loads and evaluate network performance under demanding conditions~\cite{3GPP.TR.36.814}, and the ``generate-at-will'' model for the MC user~\cite{chen2020age}, where the packet arrivals follow a Bernoulli distribution with the probability $p_{m}$ and occur immediately prior to the start of a minislot.
If a packet arrives, the MC user attempts to transmit in the minislot, and then the packet gets discarded irrespective of whether the transmission was successful or not.
The packet size of the MC is fixed and equal to $\xi$~bit, and its transmission rate is $v_{m} = \xi / \tau_\text{ms} $, where $\tau_\text{ms}$ is the minislot duration.
Both eMBB and MC users experience Rayleigh fading channels with their channel gains on allocated minislots being independent and identically distributed (i.i.d).
Let $g_{i}$, $i\in \left\{m, e \right\}$, be the channel gains of MC and eMBB users.
As $g_{i}$ follows a Rayleigh distribution, $\left | g_{i} \right |^{2}$ has an exponential distribution, with the probability density function (PDF) $p\left\{ \left | g_{i} \right |^{2} > G \right\} = \exp({-G/\mu_{ g_{i}}})$, where $\mu_{g{i} } = E_{B}\left\{\left| g_{i} \right |^{2} \right\}$.

In the considered system model, we investigate three access schemes: puncturing, NOMA, and RSMA, as shown in Fig.~\ref{fig:enter-label}(c), Fig.~\ref{fig:enter-label}(d), and Fig.~\ref{fig:enter-label}(e), respectively. The specific details of each scheme are elaborated on below.

\subsection{Puncturing}

In this approach, the eMBB user is preempted if the MC user becomes active\footnote{In this case, the BS signals a preemption indicator to the eMBB user through a physical downlink control channel, informing it to stop sending on that minislot~\cite{3GPP.TS.38.213}}.
Consequently, the transmission of the MC user in the minislot is interference-free.
The received signal at the BS, when the MC user is active in a minislot $s\in S$, can be expressed as
\begin{align*}
y = \sqrt{P_{m} d_{m}^{-\alpha}} g_m x_{m} + n,
\end{align*}
where $x_{m}$ represents the transmitted signal from the MC user, $P_{m}$ is power allocated to the MC user, $\alpha$ is the path loss exponent, and $n = \mathcal{CN}(0,\sigma^{2})$ is the complex Gaussian noise with zero mean and variance $\sigma^2$.
Thus, the SNR of the MC user at the BS is calculated as
\begin{align}
\gamma_{m} = \frac{P_m d_m^{-\alpha}\left | g_{m} \right |^{2}}{\sigma^{2} }.
\label{eq.SNR.Ru.Punc}
\end{align}

If the MC user is inactive, the eMBB user's transmission is interference-free, and the signal received at the BS is
\begin{align}
\label{eq:embb}
y = \sqrt{P_{e} d_{e}^{-\alpha}} g_e x_{e} + n,
\end{align}
where $x_{e}$ denotes the signal transmitted by the eMBB and $P_{e}$ is power allocated to the eMBB user.
In this case, the SNR at the BS is
\begin{align}
\gamma_{e} = \frac{P_e d_e^{-\alpha}\left | g_{e} \right |^{2}}{\sigma^{2} }.
\label{eq.SNR.Re.Punc}
\end{align}

\subsection{NOMA}

In NOMA, determining the decoding sequence is crucial for optimizing communication efficiency. As noted in~\cite{popovski20185g}, the decoding sequence is closely tied to the stringent latency constraints in the system. Specifically, in NOMA setups, the decoding of MC user transmissions cannot be delayed or dependent upon the decoding of eMBB traffic. This means that traditional SIC decoders, which prioritize the user with the stronger channel gain for decoding regardless of their use case are unsuitable.

In NOMA-based access, if the MC user is active, and interferes with the eMBB user, the BS attempts decoding the MC user first. If the MC packet is successfully decoded, it is removed from the received composite signal via SIC before decoding the eMBB user. On the other hand, if the MC decoding fails, SIC is not performed, and the decoding attempt of the eMBB user is subject to interference from the MC user's transmission.

The received signal at the BS when the MC user is active in minislot $s\in\mathcal{S}$ can be expressed as
\begin{align*}
 y  = \sqrt{P_{m} d_{m}^{-\alpha}} g_m x_{m}+\sqrt{P_{e} d_{e}^{-\alpha}} g_e x_{e} + n.
\end{align*}
Consequently, the signal-to-interference-plus-noise ratio (SINR) of the MC user at the BS is given by
\begin{align}
    \gamma_{m} = \frac{P_{m} d_{m}^{-\alpha}\left | g_{m} \right |^{2}}{P_e d_{e}^{-\alpha}\left | g_{e} \right |^{2} + \sigma^{2} }.
    \label{eq.SINR.Ru.NOMA}
\end{align}
Upon successful decoding of the MC user, the SNR of the eMBB user received at the BS is:   
\begin{align}
    \gamma_{e} = \frac{P_{e} d_{e}^{-\alpha}\left | g_{e} \right |^{2}}{\sigma^{2} }.
    \label{eq.SNR.Re.NOMA}
\end{align}
If the decoding of the MC user fails, the SINR of the eMBB user at the BS is
\begin{align}
    \gamma_{e} = \frac{P_{e} d_{e}^{-\alpha}\left | g_{e} \right |^{2}}{P_m d_{m}^{-\alpha}\left | g_{m} \right |^{2} + \sigma^{2} }.
    \label{eq.SINR.Re.NOMA}
\end{align}

Finally, if the MC user is inactive, then \eqref{eq:embb} and \eqref{eq.SNR.Re.Punc} hold.

\subsection{RSMA}
\label{subsec:RSMA_SM}
RSMA, initially proposed by Rimoldi in 1996~\cite{rimoldi1996rate} for the Single-Input Single-Output (SISO) Multiple Access Channel (MAC), presents an efficient solution to achieving the capacity region of the Gaussian MAC without requiring time sharing among users. In the context of RSMA, achieving the optimal rate region is feasible by defining $2M-1$ virtual users, where $M$ represents the number of users in the system~\cite{rimoldi1996rate}.
In our system model with two users (i.e., one MC and one eMBB), it is necessary to define three virtual users.
Therefore, we assume the partitioning of the MC user message into two virtual users/parts as
\begin{equation*}
    x_{m} = x_{m,1} + x_{m,2}.
\end{equation*}
Fig.~\ref{fig:enter-label}(e) illustrates the assumed scenario, where two virtual MC users with a power splitting factor $\omega \in \left ( 0,1 \right )$ and one eMBB user transmit information to the BS.

\begin{figure}[!t]
\centering{\includegraphics[width= 0.7\linewidth]{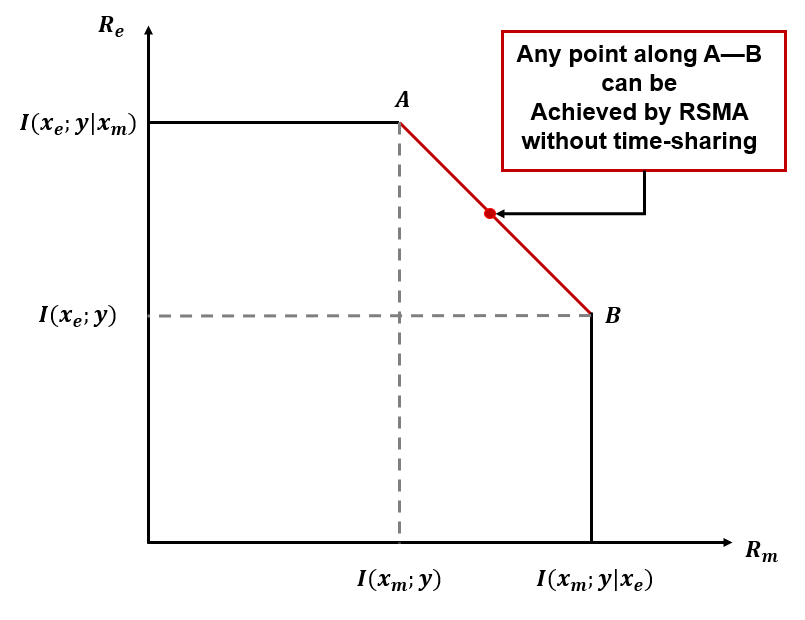}}
\vspace{-8pt}
\caption{Two-user Gaussian multiple-access capacity region. $x_{m}$ and $x_{e}$ are
the information symbols of MC and eMBB, respectively. $y$ is the received
signal. $I(x_{e};y)$ is the mutual information between $x_{e}$ and $y$. $I(x_{e};y|x_{m})$ is
the conditional mutual information between $x_{e}$ and $y$ given $x_{m}$~\cite{tse2005fundamentals,mao2022rate}.
}
\label{fig: Rate : Region}
\vspace{-10pt}
\end{figure}

 In the two-user SISO Gaussian MAC, illustrated in Fig.~\ref{fig: Rate : Region}, achieving corner points A and B involves SIC with two reverse decoding orders. However, this approach does not extend to the rate points along the A–B line segment. RSMA employing the decoding order $x_{m,1}\rightarrow x_{m,2} \rightarrow x_{e}$, achieves point A, similar to conventional NOMA with the decoding order $x_{m} \rightarrow x_{e}$. Conversely, the decoding order $x_{e} \rightarrow x_{m,1} \rightarrow x_{m,2}$ solely attains point B similar to conventional NOMA with the decoding order $x_{e} \rightarrow x_{m}$. While three methods, namely joint encoding/decoding, time sharing, and rate splitting, can achieve points along the A–B line segment, the former two methods are impractical due to their complexities and communication overhead~\cite{rimoldi1996rate}. In contrast, RSMA emerges efficiently reaches every point of the capacity region, including the A–B line segment, through SIC~\cite{mao2022rate}.
\\
Meanwhile, studies exploring the uplink scenario in RSMA have shown improvements in outage performance, which directly translates to a higher success probability for MC users and improved age performance overall~\cite{liu2020rate}. Furthermore, the work in~\cite{yang2020sum} highlights that setting up the power allocation and decoding order to maximize the total rate while meeting proportional rate constraints results in optimized rate performance, thus benefiting both the eMBB and MC users in terms of rate and age performance. Consequently, this paper adopts the optimal decoding order for uplink RSMA—starting with $x_{m,1}$, followed by $x_{e}$, and concluding with $x_{m,2}$—which has been demonstrated to effectively maximize the sum-rate, improve outage performance, and enhance AoI performance.

Assuming this decoding order, 
if $x_{m,1}$ is successfully decoded,
it is removed from the received composite signal via SIC, and then the BS attempts decoding of the eMBB user.
If $x_{e}$ is successfully decoded, SIC is again performed and the decoding attempt of $x_{m,2}$ is subject to zero interference from $x_{m,1}$ and $x_{e}$.
If SIC decoding fails for $x_{m,1}$, SIC is not performed and the decoding attempt of $x_{e}$ is subject to the complete interference of $x_{m}$, i.e., of both virtual users that comprise it.

The received signal when the MC user is active is   
\begin{align}
    y  = & \sqrt{\omega P_{m} d_{m}^{-\alpha}} g_m x_{m,1}+\sqrt{\left (1-\omega \right ) P_{m} d_{m}^{-\alpha}} g_m x_{m,2} \nonumber \\ 
    & + \sqrt{P_{e} d_{e}^{-\alpha}} g_e x_{e}  +n. 
\end{align}
The receiver first attempts decoding of the virtual MC user 1 (i.e., $x_{m,1}$) while considering all other messages as interference;
consequently, the SINR for $x_{m,1}$ is given by
\begin{align}
    \gamma_{m,1} = \frac{\omega P_{m} d_{m}^{-\alpha}\left | g_{m} \right |^{2}}{\left ( 1-\omega \right )P_m d_m^{-\alpha}\left | g_{m} \right |^{2} + P_e d_{e}^{-\alpha}\left | g_{e} \right |^{2} + \sigma^{2} }.
    \label{eq.16.SINR.Ru.1.RSMA}
\end{align}
If the decoding is successful, $x_{m,1}$ is canceled from the received signal and the SINR of the eMBB user becomes
\begin{align}
 \gamma_{e} = \frac{P_e d_{e}^{-\alpha}\left | g_{e} \right |^{2} }{\left ( 1-\omega \right )P_m d_m^{-\alpha}\left | g_{m} \right |^{2} + \sigma^{2} }.
 \label{eq.SINR.Re.no-interference.RSMA}
\end{align}
If the decoder fails in the first step, the eMBB user suffers interference from the MC user and its SINR is given by 
\begin{align}
 \gamma_{e} = \frac{P_e d_{e}^{-\alpha}\left | g_{e} \right |^{2} }{P_m d_m^{-\alpha}\left | g_{m} \right |^{2} + \sigma^{2} }.
 \label{eq.SINR.Re.interference.RSMA}
\end{align}
Finally, the SNR for the remaining virtual user $x_{m,2}$, if the decoding of the preceding users was successful, is given by
\begin{align}
\gamma_{m,2} = \frac{\left ( 1-\omega \right )P_m d_m^{-\alpha}\left | g_{m} \right |^{2}}{\sigma^{2} }.
\label{eq.SINR.Ru.2.RSMA.no-interference}
\end{align}

\subsection{Performance Metrics}

We are interested in the following performance metrics.

\subsubsection{\textbf{Successful decoding probability of the MC user}} Denoted by $\SA$, it is defined as the probability that the achievable data rate $R_m$ supported by the channel can accommodate the desired transmission data rate $v_m$, i.e.
\begin{align}
\SA = p \{ v_{m} \leq R_m \} .
\label{eq.1.success.MC}
\end{align} 
$\SA$ is a function of the SNR or SINR of the MC user (depending on the access scheme), and we compute it in Sec.~\ref{Analysis}.
We also define the outage probability of the MC user, denoted as $p_{m}^{\text{out}}$, which is simply
\begin{align}
p_{m}^{\text{out}} = p \{ v_{m} > R_m \} = 1-\SA.
\label{eq.1.success.MC}
\end{align} 

\subsubsection{\textbf{Average AoI}} AoI measures the freshness of information for time-sensitive applications and is defined as the time interval elapsed since the moment of the generation of the last successfully received packet.
In this work, we measure this time interval in minislots
\begin{figure}[!t]
\centering{\includegraphics[width= 1\linewidth]{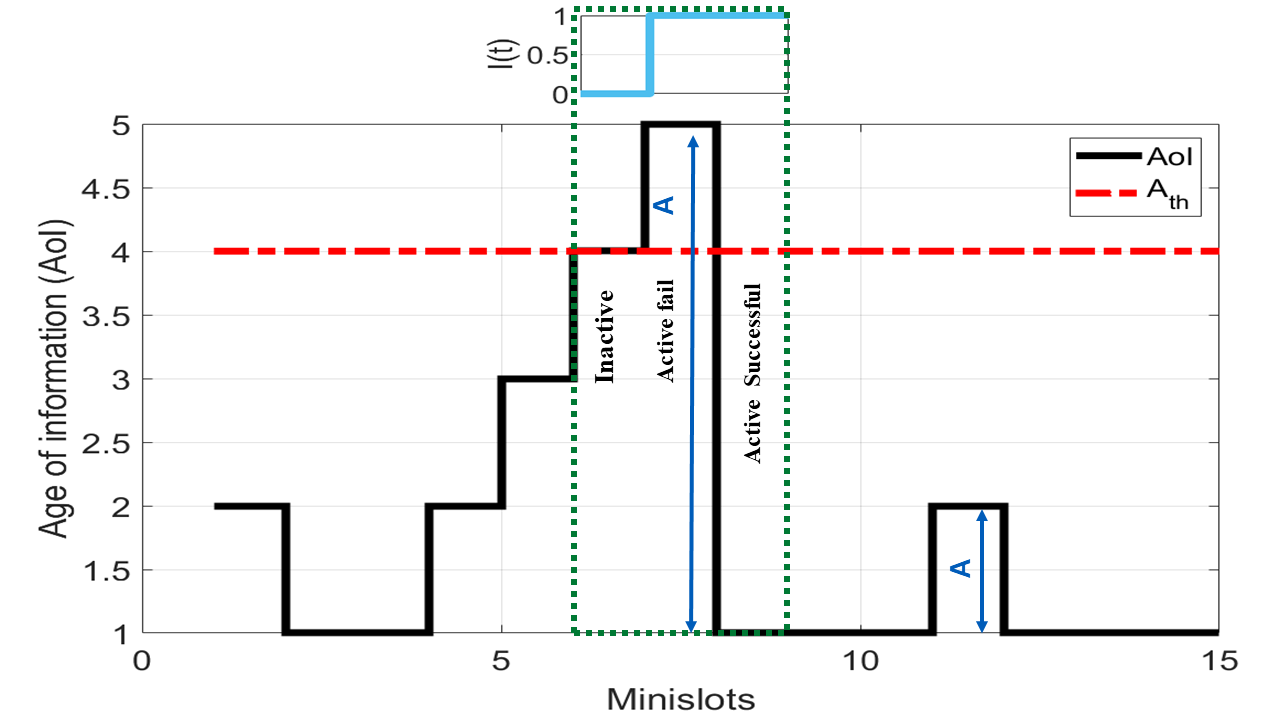}}
\caption{An example of the AoI evolution for the MC user.}
\label{fig: AoI.Model}
\vspace{-10pt}
\end{figure}
\begin{equation*}
   \Delta \left ( t+1 \right ) = \left\{\begin{matrix}
 \Delta \left ( t \right ) + 1& \left ( I\left ( t \right ) = 0 \right ) \,\,\, \mathrm{or} \\
 & \left (X\left(t\right) = 0\,\,\,\mathrm{and}\,\,\, I\left ( t \right ) = 1\right )\\[0.5em] 
 1 & \left ( I\left ( t \right ) = 1\,\,\,\mathrm{and}\,\,\, X\left(t\right) = 1 \right )
\end{matrix}\right.
\label{eq.1}
\end{equation*}
where $t$ denotes the minislot index.
The indicator function $I(t)$ is equal to 1 if the MC user experiences traffic arrival in the minislot $t$ and to 0 otherwise, while the indicator function $X(t)$ is equal to 1 if the packet is successfully received in the minislot $t$ and to 0 otherwise.
The evolution of the (instantaneous) AoI for the MC user is illustrated in Fig.~\ref{fig: AoI.Model}.
In the paper, we are interested in the average AoI $\bar{\Delta } = \mathrm{E} [\Delta ( t ) ] $.

\subsubsection{\textbf{PAoI violation probability}}
It defines the probability that the PAoI, denoted by symbol $A$, exceeds a threshold $A_{\text{th}}$, i.e.
\begin{equation*}
     p_v =  p\left \{ A > A_{\text{th}} \right \},
\end{equation*} 
Fig.~\ref{fig: AoI.Model} shows one instance of PAoI that exceeds and one that does not exceed a threshold $A_{\text{th}} = 4$ (dotted red line).
\subsubsection{\textbf{Data rate of the eMBB user}}
Denoted by $R_e$, it is a function of the SNR/SINR, depending on the access method and the activity of the MC user in the case of NOMA and RSMA.

In the next section, we analytically derive these performance parameters for the considered transmission strategies.

\section{Analytical Modeling and Derivations}
\label{Analysis}



\begin{figure}[!t]
\centering{\includegraphics[width=0.95\linewidth]{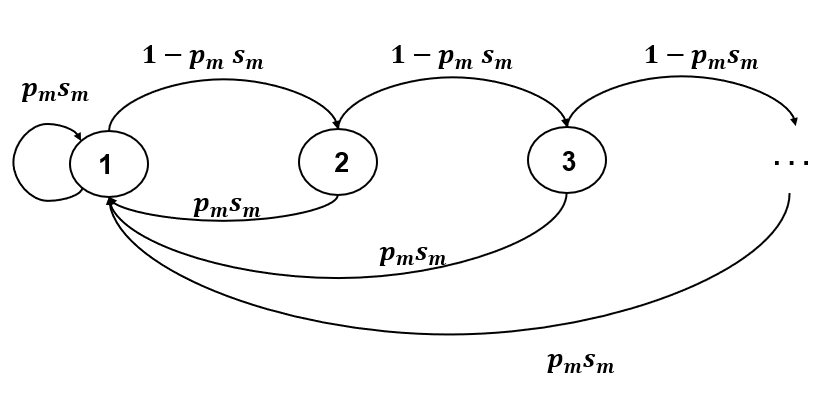}}
\caption{The DTMC model representing the AoI evolution of MC traffic.}
\label{fig:DTMC.Model}
\end{figure}

We start by deriving the general formulas for the average AoI and the PAoI violation probability for the MC user.
Fig.~\ref{fig:DTMC.Model} depicts the DTMC model of the AoI evolution for the MC user, where each state represents the instantaneous value of AoI.
The DTMC transits from state $j$ to state $j+1$ when the MC transmission is unsuccessful or if the MC user is inactive.
On the other hand, the DTMC transits from state $j$ to state 1 if the MC transmission is successful.
The transition matrix of this DMTC can be written as
\begin{align}
\nonumber
\mathbf{P}_{\text{AoI}}=\begin{bmatrix}
 \PMC \SA & 1-\PMC \SA & 0 & 0 & \cdots \\ 
\PMC  \SA & 0 & 1-\PMC  \SA & 0 & \cdots\\ 
 \PMC  \SA & 0 & 0 &  1- \PMC  \SA & \cdots \\ 
\vdots & \vdots & \vdots & \vdots &  \ddots   
\end{bmatrix}
\label{eq.3}
\end{align}
where $\PMC$ is the activation probability of the MC user, see Sec.~\ref{System.Model}.

To calculate the average AoI, we need to calculate the steady-state probability matrix $\boldsymbol{\pi} _{\text{AoI}}=\left [\pi _{1},\pi _{2},...,\pi _{l},...  \right ]$.
The probability that the AoI is equal to $l$ at the steady state is defined as $\pi _{l}=  \PMC \SA \left (1- \PMC \SA \right )^{l-1}$, $\forall l$.
By using the set of equations $\boldsymbol{\pi} _{\text{AoI}} \mathbf{P}_{\text{AoI}}= \boldsymbol{\pi} _{\text{AoI}} $ and $\sum_{l=1}^{\infty } \pi _{l}=1$, the average AoI can be computed as 
\begin{equation*}
\overline{\Delta }=\sum_{l=1}^{\infty }\pi _{l}  l =  \PMC \SA  \sum_{l=1}^{\infty } l \left (1- \PMC \SA  \right )^{l-1}. 
\label{eq.5}
\end{equation*}
Since $ 1- \PMC \SA  < 1 $, the previous equation reduces to 
\begin{equation}
\overline{\Delta}= \frac{p_{m}s_{m}}{\left ( 1-\left (1-p_{m}s_{m}\right ) \right )^{2}} =  \frac{1}{p_{m}s_{m}}.
\label{eq.4.average.AoI}
\end{equation}

To calculate the PAoI violation probability, we determine the PAoI distribution and the probability of exceeding a threshold.
In our system model, where packet arrivals follow a Bernoulli distribution with probability $p_m$, the resulting PAoI distribution is characterized by a geometric distribution.
Specifically, the geometric distribution provides the probability that the first success occurs after exactly $A$ independent trials, where each has a success probability of $p_{m}s_{m}$.
If the likelihood of success on each trial is $p_{m}s_{m}$, then the cumulative mass function (CMF) of geometric distribution for PAoI is expressed as 
\begin{align}
    p\left \{ A \leq A_{\text{th}} \right \} =\left\{\begin{matrix}
 1-\left ( 1-p_{m}s_{m} \right )^{\left \lfloor A_{\text{th}} \right \rfloor}& A_{\text{th}} \geq 1 \\ 
 0 & A_{\text{th}} < 1
\end{matrix}\right.
\label{eq.new.num}
\end{align}
Using~\eqref{eq.new.num}, the PAoI violation probability becomes 
\begin{equation}
    p_{v} = p\left \{ A > A_{\text{th}} \right \} = 1- p\left \{ A \leq A_{\text{th}} \right \} = \left ( 1-p_{m}s_{m} \right )^{\left \lfloor A_{\text{th}} \right \rfloor}.
    \label{eq.6.PAoI}
\end{equation}

In what follows, we analyze the considered access schemes, where we derive the data rate $R_e$ for the eMBB user and $R_m$ for MC user, as well as derive $s_m$ for each access scheme to obtain $\overline{\Delta}$ and $p_v$, according to \eqref{eq.4.average.AoI} and \eqref{eq.new.num}.

\subsection{Puncturing}
If the MC user is active, the eMBB transmission is preempted by the BS.
Thus, leveraging \eqref{eq.SNR.Ru.Punc}, the achievable data rate for the MC user can be determined as
\begin{align*}
 R_{m}^{\text{pun}} = \frac{BW}{S} \log_{2}\left ( 1+\frac{P_{m} \left | g_{m} \right |^{2} d_{m}^{-\alpha}}{\sigma ^{2}} \right),
\end{align*}
where $BW / S$ is the bandwidth of a minislot and $\left | g_{m} \right |^{2}$ is the average channel gain in each resource block for MC transmission.
The success probability of MC transmission is then
\begin{align}
s_{m}^{\text{punc}} = &
p\left \{ R_{m}^{\text{punc}} > v_{m}\right \} =
p\left \{ \left | g_{m} \right |^{2} > \frac{\rho \sigma ^{2}}{ P_{m} d_{m}^{-\alpha}}\right \}  \nonumber \\
= & \exp\left({-\frac{\rho \sigma ^{2}}{P_{m} d_{m}^{-\alpha} \mu _{ g_{m}}}}\right),
\label{eq.3.su.punc}
\end{align}
where $\rho = 2^{\frac{v_{m} S}{BW}}-1$.\\


By substituting \eqref{eq.3.su.punc} into \eqref{eq.4.average.AoI}, the average AoI becomes 
\begin{equation}
 \overline{\Delta}_\text{punc} = \frac{1}{p_{m}} \exp\left({\frac{\rho \sigma ^{2}}{P_{m} d_{m}^{-\alpha} \mu _{g_{m}}}}\right).
 \label{eq.5.AAoI.Punc}
\end{equation}
Further, by substituting \eqref{eq.3.su.punc} into \eqref{eq.6.PAoI}, the PAoI violation probability becomes
\begin{align}
    p_{v}^{\text{punc}} = \left ( 1-p_{m}\cdot\exp\left({-\frac{\rho \sigma ^{2}}{P_{m} d_{m}^{-\alpha} \mu _{ g_{m}}}}\right)\right )^{\left \lfloor A_{\text{th}} \right \rfloor}.
    \label{eq.7.PAoI.punc}
\end{align}

Finally, since the eMBB user does not suffer from interference in this approach, its data rate can be determined using~\eqref{eq.SNR.Re.Punc}
\begin{align}
 R_{\text{e}}^{\text{punc}} = BW \left ( 1- \frac{\chi}{S} \right ) \log_{2}\left ( 1+\frac{P_{e} \left | g_{e} \right |^{2}d_{e}^{-\alpha}}{\sigma ^{2}} \right ),
 \label{eq.8.Re.punc}
\end{align}
in this context, $\chi$ denotes the average number of minislots in which the MC user is active during each time slot. Since both $\frac{\chi}{S}$ and $p_{m}$ represent the average number of active MC users per time slot, we can replace $\frac{\chi}{S}$ with $p_{m}$. Consequently, the achievable rate for the eMBB user in the puncturing scheme can be defined as 
\begin{align}
 R_{\text{e}}^{\text{punc}} = BW \left ( 1- p_{m} \right ) \log_{2}\left ( 1+\frac{P_{e} \left | g_{e} \right |^{2}d_{e}^{-\alpha}}{\sigma ^{2}} \right ),
 \label{eq.8.Re.punc}
\end{align}\\
where  $1- p_{m}$ is the expected fraction of the minislots in which the eMBB user is active and $\left | g_{e} \right |^{2}$ is the average channel gain in the RBs of the eMBB user. 
\subsection{NOMA}
\label{sec:NOMA_analysis}
In this case, the achievable rate of the MC user can be determined using~\eqref{eq.SINR.Ru.NOMA} as
\begin{align*}
 R_{m}^{\text{NOMA}} = \frac{BW}{S} \log_{2}\left ( 1+\frac{P_{m} \left | g_{m} \right |^{2}d_{m}^{-\alpha}}{P_{e} \left | g_{e} \right |^{2} d_{e}^{-\alpha} + \sigma ^{2}} \right ).
\end{align*}
In addition, the probability of successful MC transmission is 
\begin{align}
s_{m}^{\text{NOMA}} & = p\left \{ R_{m}^{\text{NOMA}} > v_{m}\right \} \nonumber \\
& = p\left \{ \left | g_{m} \right |^{2} > \frac{\rho \left [ P_{e} \left | g_{e} \right |^{2} d_{e}^{-\alpha} +\sigma ^{2} \right ]}{P_{m} d_{m}^{-\alpha}}\right \},
\label{eq.10.su.NOMA.initial}
\end{align}
which can be calculated as  
\begin{align}
    & s_{m}^{\text{NOMA}} = 
    \int_{z=0}^{\infty }f_{z}(Z)  p\left \{ \left | g_{m} \right |^{2} > \frac{\rho \left [ P_{e} d_{e}^{-\alpha} z +\sigma ^{2} \right ]}{P_{m} d_{m}^{-\alpha}}\right \} dz \nonumber \\
    & = \int_{z=0}^{\infty}  \frac{1}{\mu _{g_{e} }} \exp \left(-\frac{z}{\mu_{g_{e}}} \right) \exp \left(-\frac{\rho \left [ P_{e} d_{e}^{-\alpha} z+ \sigma ^{2}\right ]}{P_{m}d_{m}^{-\alpha}\mu _{g_{m}}} \right) dz \nonumber\\
    &   = \frac{P_{m} d_{m}^{-\alpha} \mu _{g_{m}}}{P_{m} d_{m}^{-\alpha} \mu _{g_{m}} + \rho P_{e} d_{e}^{-\alpha} \mu _{ g_{e}} }  \exp\left(-\frac{\rho \sigma ^{2}}{P_{m} d_{m}^{-\alpha} \mu _{g_{m}}}\right).
    \label{eq.11.su.NOMA}
\end{align}
By substituting \eqref{eq.11.su.NOMA} into \eqref{eq.4.average.AoI}, the average AoI becomes 
\begin{equation}
\overline{\Delta}_{\text{NOMA}}  = \frac{P_{m} d_{m}^{-\alpha}  \mu _{g_{m} } + \rho P_{e} d_{e}^{-\alpha} \mu _{ g_{e}}}{p_{m} \cdot \left[P_{m} d_{m}^{-\alpha} \mu _{g_{m}}\right ]} \exp\left({\frac{\rho \sigma ^{2}}{P_{m} d_{m}^{-\alpha}  \mu _{g_{m} }}}\right).
    \label{eq.12.AAoI.NOMA}
\end{equation}
Meanwhile, the PAoI violation probability can be derived by substituting \eqref{eq.11.su.NOMA} into \eqref{eq.6.PAoI}
\begin{align}
     & 
     p_{v}^{\text{NOMA}}  = 
     \nonumber \\ 
     &
     \left ( 1-\frac{p_{m}P_{m} d_{m}^{-\alpha} \mu _{g_{m}}}{P_{m} d_{m}^{-\alpha} \mu _{g_{m}} + \rho P_{e} d_{e}^{-\alpha} \mu _{ g_{e}} }  \exp\left(-\frac{\rho \sigma ^{2}}{P_{m} d_{m}^{-\alpha} \mu _{g_{m}}}\right)\right )^{\!\!\left \lfloor A_{\text{th}} \right \rfloor}\!\!\!.
     \label{eq.13.PAoI.NOMA}
\end{align}

In the NOMA scenario, the achievable rate for the eMBB user depends on the activation probability of the MC user, and the number of successful and unsuccessful decodings of the interfering MC transmissions (i.e.,  when the MC user is active).
The expected fraction of the minislots in which the MC user is active is simply $p_m$; this fraction can be further subdivided in  $s_{m}^{\text{NOMA}} p_m$ subfraction when the MC user is successfully decoded, causing no interference to the eMBB user, and $(1 - s_{m}^{\text{NOMA}}) p_m$ when it is not decoded successfully.
Thus, using~\eqref{eq.SNR.Re.NOMA} and \eqref{eq.SINR.Re.NOMA}, the achievable rate for the eMBB user can be expressed as
\begin{align}
 & R_{\text{e}}^{\text{NOMA}}  = BW \left ( 1- p_{m} \right ) \log_{2}\left ( 1+\frac{P_{\text{e}} \left | g_{\text{e}} \right |^{2}d_{e}^{-\alpha}}{\sigma ^{2}} \right ) \nonumber \\
 & + BW p_{m} (1 - s_{m}^{\text{NOMA}}) \log_{2}\left ( 1+\frac{P_{\text{e}} \left | g_{\text{e}} \right |^{2}d_{e}^{-\alpha}}{P_{m} \left | g_{m} \right |^{2} d_{m}^{-\alpha} + \sigma ^{2}} \right ) \nonumber \\
 & + BW p_{m} s_{m}^{\text{NOMA}} \log_{2}\left ( 1+\frac{P_{\text{e}} \left | g_{\text{e}} \right |^{2}d_{e}^{-\alpha}}{\sigma ^{2}} \right ).
 \label{eq.13.Re.NOMA}
\end{align}

\subsection{RSMA}

The target data rates for transmitting MC virtual users $x_{m,1}$ and $x_{m,2}$ can be defined as $\nu _{m,1} = \lambda \nu _{m}$ and $\nu _{m,2} = \left ( 1-\lambda \right ) \nu _{m}$, respectively, where $\lambda \in (0,1)$ represents the target data rate factor.
For the selected decoding order (c.f., Sec.~\ref{subsec:RSMA_SM}), the MC user would encounter an outage if any of the following three conditions occur:
\begin{itemize}
    \item $x_{m,1}$ is not successfully decoded.
    \item $x_{m,1}$ is successfully decoded but $x_{e}$ is not successfully decoded. 
    \item $x_{m,1}$ and $x_{e}$ are successfully decoded, but $x_{m,2}$ is not successfully decoded.
\end{itemize}
Then, the outage probability of the MC user can be determined as 
\begin{align}
    p_{m}^{\text{out}} = & 
    p\left \{ R_{m,1}< \lambda \nu _{m} \right \}+p\left \{ R_{m,1}\geq \lambda \nu _{m} \right \} p\left \{ R_{e}< \nu_{e} \right \} \nonumber \\ +
    & p\left \{ R_{m,1} \geq \lambda \nu _{m} \right \}  p\left \{ R_{e} \geq \nu_{e} \right \}   p\left \{ R_{m,2} < (1-\lambda) \nu_{m} \right \},
     \label{eq.15.pout.initial.MC.RSMA}
\end{align}
where $\nu_{e}$ is the target data rate for the eMBB user. 

In our study, we ensure consistency between NOMA and RSMA schemes by adopting a comparable approach regarding data rate targets. In NOMA, no specific data rate target is assumed, and we extend this principle to RSMA. Our objective in RSMA is to maximize the data rate for users without specifying a target rate in order to comparing the performance of NOMA and RSMA under similar conditions. Consequently, the eMBB user can be assumed to consistently satisfy the conditions $p\{R_{e} < \nu_{e}\} = 0$ and $p\{R_{e} \geq \nu_{e}\} = 1$.
Thus, \eqref{eq.15.pout.initial.MC.RSMA} simplifies into
\begin{align}
 p_{m}^{\text{out}} & =  p\left \{ R_{m,1}< \lambda \nu_{m} \right \} \nonumber \\  & + p\left \{ R_{m,1} \geq \lambda \nu_{m} \right \}  p\left \{ R_{m,2} < (1-\lambda) \nu_{m} \right \}.
    \label{eq.15.pout.MC.RSMA}
\end{align}
Then, the probability of successful transmission for the MC user is simply 
\begin{align}
 s_{m}^{\text{RSMA}} = 1-p_{m}^{\text{out}}. 
\label{eq.13.su.RSMA}   
\end{align}



To obtain the outage probability in \eqref{eq.15.pout.MC.RSMA}, we begin by deriving a closed form for $p\left \{ R_{m,1}< \lambda \nu_{m} \right \}$ in the initial decoding step
\begin{align}
    p\left \{ R_{m,1}< \lambda \nu_{m}  \right \} & = p\left \{ \frac{BW}{S} \log_{2} \left ( 1+\gamma_{m,1} \right )< \lambda \nu_{m}  \right \} \nonumber \\ & =
     p\left \{ \gamma_{m,1} < 2^{\frac{\lambda \cdot \nu_{m} \cdot S}{BW}}-1\right \} \nonumber \\ & =  p\left \{ \gamma_{m,1} < \rho_{1} \right \}.
     \label{eq.18.p.out.1.initial}
\end{align}
By substituting \eqref{eq.16.SINR.Ru.1.RSMA} into \eqref{eq.18.p.out.1.initial}, we have
\begin{align}
 p\left \{ R_{m,1}< \lambda \nu_{m} \right \} & = p\left \{ \left | g_{m} \right | ^{2} < \frac{\rho_{1}  \left ( P_e  d_e^{-\alpha} \left | g_{e} \right | ^{2}+\sigma ^{2} \right )}{P_m  d_m^{-\alpha} \left (\omega -\rho_{1} \left ( 1-\omega \right )\right )}\right \}  . 
 \label{eq.19.p.out.1.initial.2}
\end{align}
Since $\omega - \rho_{1} \left( 1- \omega \right) $ is always positive, the outage probability for $x_{m,1}$ is 
\begin{align}
    & p\left \{ R_{m,1}< \lambda \nu_{m}  \right \} \nonumber \\ & = 
     \int_{z=0}^{\infty }f_{z}(Z)  p\left \{ \left | g_{m} \right | ^{2} < \frac{\rho_{1} \left ( P_e  d_e^{-\alpha} z +\sigma ^{2} \right )}{P_m  d_m^{-\alpha} \left (\omega -\rho_{1} \left ( 1-\omega \right )\right )}\right \}  dz   \nonumber \\
    & = 1-\frac{P_{m} d_{m}^{-\alpha} \left(\omega - \rho_{1} (1-\omega)\right)}{P_{m} d_{m}^{-\alpha}  \left(\omega - \rho_{1} (1-\omega)\right) + \rho_{1} P_{e} d_{e}^{-\alpha} } \nonumber \\
    & \times \exp{\left(-\frac{\rho_{1} \sigma ^{2}}{P_{m} d_{m}^{-\alpha} \mu _{g_{m}}\left(\omega - \rho_{1} (1-\omega)\right)}\right)}.
    \label{eq.20.p.out.1}
\end{align}
The success probability of the initial segment of the MC user, denoted as $p\left \{ R_{m,1} \geq \lambda \nu_{m} \right \}$, is simply the complement of the outage probability derived in \eqref{eq.20.p.out.1}
\begin{align}
  & p\left \{ R_{m,1} \geq \lambda \nu_{m}  \right \}  =  1- p\left \{ R_{m,1} < \lambda \nu_{m}  \right \} \nonumber \\
 & = \frac{P_{m} d_{m}^{-\alpha} \left(\omega - \rho_{1} (1-\omega)\right)}{P_{m} d_{m}^{-\alpha} \left(\omega - \rho_{1} (1-\omega)\right) + \rho_{1} P_{e} d_{e}^{-\alpha} } \nonumber \\
    & \times \exp{\left(-\frac{\rho_{1} \sigma ^{2}}{P_{m} d_{m}^{-\alpha} \mu _{g_{m}}\left(\omega - \rho_{1} (1-\omega)\right)}\right)}.
  \label{eq.22.s.u.1}
\end{align}
Finally, the outage probability of the MC user in the third step of decoding can be computed as:
\begin{align}
    & p\left \{ R_{m,2}< (1-\lambda) \nu_{m}  \right \} \nonumber \\ =
    & p\left \{ \frac{BW}{S} \log_{2} \left ( 1+\gamma_{2}^{u} \right )< (1-\lambda) \nu_{m}  \right \} \nonumber \\ =
     & p\left \{ \gamma _{2}^{u} < 2^{\frac{(1-\lambda)\nu_{m} \cdot S}{BW}}-1\right \} =  p\left \{ \gamma _{2}^{u} < \rho_{2} \right \}.
     \label{eq.23.p.out.2.initial}
\end{align}
As our focus lies on the outage probability of $x_{m,2}$ concerning successful SIC decoding in prior steps, substituting \eqref{eq.SINR.Ru.2.RSMA.no-interference} into \eqref{eq.23.p.out.2.initial} yields
\begin{align}
 p\left \{ R_{m,2}< (1-\lambda) \nu_{m}  \right \}= p\left \{ \left | g_{m} \right | ^{2} < \frac{\rho_{2} \, \sigma ^{2}}{P_m  d_m^{-\alpha} \left ( 1-\omega \right )}\right \}.
 \label{eq.24.p.out.2.initial.2}
\end{align}
Since $\omega < 1 $, $p\left \{ R_{m,2}< (1-\lambda) \nu_{m}  \right \}$ can be derived as 
\begin{align}
     p\left \{ R_{m,2}< (1-\lambda) \nu_{m}  \right \} = 1-\exp{\left(-\frac{\rho_{2} \sigma ^{2}}{P_{m} d_{m}^{-\alpha} \mu _{g_{m}}(1-\omega)}\right)}.
     \label{eq.25.p.out.2}
\end{align}
By substituting \eqref{eq.20.p.out.1}, \eqref{eq.22.s.u.1} and \eqref{eq.25.p.out.2} into \eqref{eq.15.pout.MC.RSMA}, the outage probability for MC user can be calculated as
\begin{align}
& p_{m}^{\text{out}} = 1 - \frac{P_{m} d_{m}^{-\alpha}   \left(\omega - \rho_{1} (1-\omega)\right)}{P_{m} d_{m}^{-\alpha}  \left(\omega - \rho_{1} (1-\omega)\right) + \rho_{1} P_{e} d_{e}^{-\alpha}} \nonumber \\
& \times \exp{\left(-\frac{(1-\omega) \rho_{1} \sigma ^{2}+\left(\omega - \rho_{1} (1-\omega)\right) \rho_{2} \sigma ^{2}}{P_{m} d_{m}^{-\alpha} \mu _{g_{m}} \left(1-\omega\right) \left(\omega - \rho_{1} (1-\omega)\right)}\right)}.
\label{p.out.MC.last}
\end{align}

Finally, substituting \eqref{p.out.MC.last} into \eqref{eq.13.su.RSMA} the success probability of MC user is derived as
\begin{align}
& s_{m}^{\text{RSMA}} = \frac{P_{m} d_{m}^{-\alpha}  \left(\omega - \rho_{1} (1-\omega)\right)}{P_{m} d_{m}^{-\alpha} \left(\omega - \rho_{1} (1-\omega)\right) + \rho_{1} P_{e} d_{e}^{-\alpha} } \nonumber \\
& \times \exp{\left(-\frac{(1-\omega) \rho_{1} \sigma ^{2}+\left(\omega - \rho_{1} (1-\omega)\right) \rho_{2} \sigma ^{2}}{P_{m} d_{m}^{-\alpha} \mu _{g_{m}} \left(1-\omega\right) \left(\omega - \rho_{1} (1-\omega)\right)}\right)}.
\label{su.MC.last}
\end{align}

Similar to puncturing and NOMA scenario, the average AoI can be derived as 
\begin{align}
       \overline{\Delta}_{\text{RSMA}} & =  \frac{P_{m} d_{m}^{-\alpha}  \left(\omega - \rho_{1} (1-\omega)\right) + \rho_{1} P_{e} d_{e}^{-\alpha} }{p_{m} \left[P_{m} d_{m}^{-\alpha}  \left(\omega - \rho_{1} (1-\omega)\right)\right]} \nonumber \\
& \times \exp{\left(\frac{(1-\omega) \rho_{1} \sigma ^{2}+\left(\omega - \rho_{1} (1-\omega)\right) \rho_{2} \sigma ^{2}}{P_{m} d_{m}^{-\alpha} \mu _{g_{m}} \left(1-\omega \right) \left(\omega - \rho_{1} (1-\omega)\right)}\right)}, 
       \label{RSMA.average}
\end{align}
and PAoI violation probability is defined as 
\begin{align}
   p_{v}^{\text{RSMA}} = \left ( 1-p_{m}s_{m}^{\text{RSMA}} \right )^{\left \lfloor A_{\text{th}} \right \rfloor}.
 \label{RSMA.PAoI}
\end{align}

In the RSMA scenario, the achievable rate for the eMBB user relies on the activity pattern of the MC user and the probability of the successful decoding of the interfering transmissions of the MC users.
Using reasoning analogous to the one in Section~\ref{sec:NOMA_analysis} for the NOMA case, the achievable rate of the eMBB user 
can be derived as
\begin{align}
& R_{\text{e}}^{\text{RSMA}}  = BW \left ( 1- p_{m} \right ) \log_{2}\left ( 1+\frac{P_{\text{e}} \left | g_{\text{e}} \right |^{2}d_{e}^{-\alpha}}{\sigma ^{2}} \right ) \nonumber \\
 & + BW p_{m}(1-s_{m,1}^{\text{RSMA}}) \log_{2}\left ( 1+\frac{P_{\text{e}} \left | g_{\text{e}} \right |^{2}d_{e}^{-\alpha}}{P_{m} \left | g_{m} \right |^{2} d_{m}^{-\alpha} + \sigma ^{2}} \right ) \nonumber \\
 & + BW p_{m} s_{m,1}^{\text{RSMA}}  \log_{2}\left ( 1+\frac{P_e d_{e}^{-\alpha}\left | g_{e} \right |^{2} }{\left ( 1-\omega^{*} \right )P_m d_m^{-\alpha}\left | g_{m} \right |^{2} + \sigma^{2} } \right ),
 \label{eq.Re.RSMA}
\end{align}
where $s_{m,1}^{\text{RSMA}} = p\left \{ R_{m,1} \geq \lambda \nu_{m} \right \}$ and $\omega^{*}$ is the optimal power allocation for RSMA to maximize the success probability of MC user. 

As evident from~\eqref{su.MC.last}, and consequently reflected in~\eqref{RSMA.average} and \eqref{RSMA.PAoI}, the success probability, average AoI, and PAoI violation probability for the MC user depends on both the power allocation ($\omega$) and rate splitting ($\lambda$) factors. 
To identify the optimal power allocation and rate splitting factor, we formulate an optimization problem aimed at maximizing the success probability of the MC user, 
expressed as
\begin{subequations}
\begin{alignat}{2}
 & \quad \underset{\omega,\lambda} {\max} \quad s_{m}^{\text{RSMA}}(\omega,\lambda)  & \label{eq.OP.RSMA.a}\\
 & \text{s.t.} \quad  0 < \lambda < 1, & \label{eq.OP.RSMA.c} \\
 & \quad \quad 0 < \omega < 1. & \label{eq.OP.RSMA.d}
\end{alignat}
\begin{figure}[!t]
\centering{\includegraphics[width=0.95\linewidth]{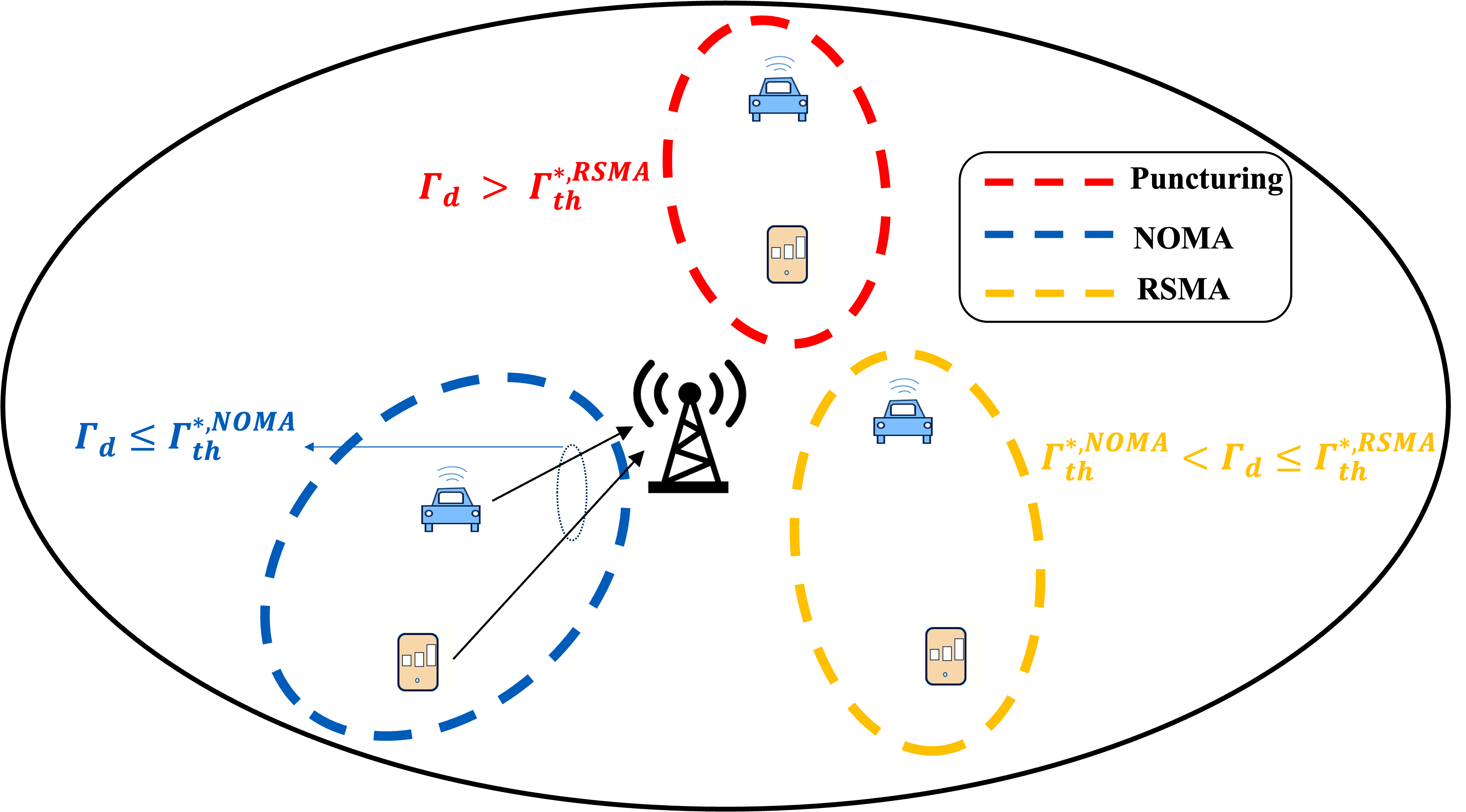}}
\vspace{-8pt}
\caption{
Illustration of optimizing the choice among puncturing, NOMA, or RSMA w.r.t. the thresholds $\Gamma_{\text{th}}^{*,\text{NOMA}}$ and $\Gamma_{\text{th}}^{*,\text{RSMA}}$.}
\label{fig:Scheduling}
\end{figure}
The objective function in~\eqref{eq.OP.RSMA.a}, with respect to $\omega$ and $\lambda$, is a non-convex and non-concave function, as the hessian matrix lacks positive or negative semi-definiteness. Consequently, the optimal values for power and rate splitting factors can be determined through an exhaustive search or the grey wolf optimizer (GWO)~\cite{faris2018grey,mirjalili2014grey}.

Unlike evaluating all possible solutions in an exhaustive search, GWO ensures a diverse search of the solution space to avoid local optimal (exploration) and focuses on the regions likely containing optimal or near-optimal solutions (exploitation) for effective optimization. The complexity of GWO depends on the number of wolves in the population ($W$), the number of iterations ($I$), and the cost of a single evaluation of the objective function. For each iteration, the algorithm evaluates the objective function for each wolf and updates their positions, resulting in a total complexity of $O(W \times I)$. In contrast, the exhaustive search method's complexity is $O(N_{\omega} \times N_{\lambda})$, where $N_{\omega}$ and $N_{\lambda}$ denote the number of discretized points in the search grid for $\omega$ and $\lambda$, respectively. Although each evaluation during this search takes constant time $O(1)$, the complexity varies depending on how finely $\omega$ and $\lambda$ are discretized.

For numerical analysis (Sec.\ref{Evaluation}), we used GWO with a population size of 50 wolves and 200 iterations, resulting in a complexity of $10^4$. In comparison, we discretized $\omega$ and $\lambda$ into $10^3$ points each for the exhaustive search, leading to a complexity of $10^6$. This demonstrates a substantial reduction in computational complexity, making GWO a practical alternative for our problem.

To address real-time challenges, a practical solution is precomputing optimal values offline based on the SNR gap ($\Gamma_{d}$) and storing them in a lookup table.

\end{subequations}
\section{Optimizing the Choice of the Transmission Strategy}
\label{Sec.Optimization}

The overall objective considered in this paper is to maximize the eMBB rate by strategically selecting the optimal access scheme (i.e., puncturing, NOMA, or RSMA) while ensuring certain guaranteed performance in terms of the average AoI and PAoI violation probability for the MC user.

Among the three strategies, puncturing performs the best in terms of the success probability of the MC user, and consequently, in average AoI and PAoI violation probability (c.f., Sec.~\ref{Evaluation}). However, puncturing has an impact on eMBB user's achievable data rate.
Therefore, while maximizing the eMBB rate, our goal is to determine the range of values of the SNR gap between the eMBB and MC users, defined as
\begin{align}
   \Gamma_{d} = 10 \log_{10}\left ( \frac{P_{e}d_{e}^{-\alpha}\left | g_{e} \right | ^{2}}{P_{m}d_{m}^{-\alpha}\left | g_{m} \right |^{2}} \right ),
\end{align}
for which NOMA/RSMA can achieve the average AoI performance and PAoI violation probability that are within a small gap compared to their best possible values attained with puncturing. 


The SNR gap, $\Gamma_{d}$ plays a crucial role in determining the optimal access scheme based solely on the SNR difference between the eMBB and MC user. $\Gamma_{d}$ is influenced by the distances of the eMBB and MC users from the BS, consequently the channel gain from the BS. Changes in the distances of eMBB and MC users alter $\Gamma_{d}$, subsequently impacting $\bar{\Delta}$, $p_{v}$, and $R_{e}$. However, the direct impact of $\Gamma_{d}$ on these metrics remains implicit. The motivation behind introducing $\Gamma_{d}$ is to account for the distances of eMBB and MC users from the BS, allowing us to investigate how distance variations influence $\bar{\Delta}$, $R_{e}$, and $p_{v}$, with the aim of clarifying the underlying dependencies and their effects on system performance.

Formally, we define two rate optimization problems, considering average AoI and PAoI violation probability as
\begin{subequations}
\begin{alignat}{2}
 & \quad \underset{i} {\max} \quad R_{e}^{i}(\Gamma_{d}) & \label{eq.OP.a.average.AoI}\\
 & \text{s.t.} \quad  \beta \geq \overline{\Delta}_{\text{i}}(\Gamma_{d}) - \overline{\Delta}_{\text{punc}}(\Gamma_{d}) & \label{eq.OP.b.average.AoI}
\end{alignat}
\end{subequations}
\begin{subequations}
\begin{alignat}{2}
 & \quad \underset{i} {\max} \quad R_{e}^{i}(\Gamma_{d}) & \label{eq.OP.a.PAoI}\\
 & \text{s.t.} \quad \hat{\beta} \geq p_{v}^{i}\left(\Gamma_{d}\right) - p_{v}^{\text{punc}} \left(\Gamma_{d}\right)   & \label{eq.OP.b.PAoI} 
\end{alignat}
\end{subequations}
where $i\in \{\text{NOMA},\text{RSMA}\}$ while $\beta$ and $\hat{\beta}$ are tolerated loss in average AoI and PAoI violation probability, respectively. 

$\Gamma_{\text{th}, \bar{\Delta}}^{i}$ and $\Gamma_{\text{th},p_v}^{i}$ are derived by first computing the values of $R_{e}$, $\bar{\Delta}$, and $p_{v}$ across various $\Gamma_{d}$ values. Later, we identify a threshold that ensures the gap between $\bar{\Delta}$ and $p_{v}$ of NOMA/RSMA with puncturing remains below $\beta$ and $\hat{\beta}$, respectively. Using closed-form expressions for $R_{e}$, $\bar{\Delta}$, and $p_{v}$ with different $\Gamma_{d}$ values \eqref{eq.5.AAoI.Punc}, \eqref{eq.7.PAoI.punc}, \eqref{eq.8.Re.punc}, \eqref{eq.12.AAoI.NOMA}, \eqref{eq.13.PAoI.NOMA}, \eqref{eq.13.Re.NOMA}, \eqref{RSMA.average}, \eqref{RSMA.PAoI} and \eqref{eq.Re.RSMA}. Through this analysis, we determine the thresholds $\Gamma_{\text{th}, \bar{\Delta}}^{i}$ and $\Gamma_{\text{th},p_v}^{i}$.

If one is interested in ensuring that both $\bar{\Delta }$ and $p_v$ simultaneously satisfy ~\eqref{eq.OP.b.average.AoI} and \eqref{eq.OP.b.PAoI}, respectively, then the optimal value of the threshold is 
\begin{equation}
    \Gamma_{\text{th}}^{*,i} = \min\left \{\Gamma_{\text{th}, \bar{\Delta } }^{i},\Gamma_{\text{th},p_v}^{i}\right \}.
    \label{threshold.optimal}
\end{equation}
Fig.~\ref{fig:Scheduling} shows an illustration of optimizing the choice of transmission strategy. 
\section{Evaluation}
\label{Evaluation}
To find a balanced medium access strategy for the eMBB-MC coexistence considered in this paper, we perform numerical- and simulation-based evaluations according to the system model in Sec.~\ref{System.Model}.
The evaluation parameters are listed in Table~\ref{table.1}.
The numerical evaluation is based on Section~\ref{Analysis} while 
the simulations are conducted by implementing the three schemes in a Matlab-based discrete-event simulator.
The obtained results are averaged over $10^5$ time slots for each set of the configuration parameters, showing a close match between the analytical and the simulation results.
\begin{table}[t]
\centering
    \caption{Evaluation Parameters~\cite{3GPP.TS23.501}.}
    \resizebox{0.48\textwidth}{!}{
    \begin{tabular}{lc}
    \toprule
     \textbf{Parameter}  & \textbf{Value}  \\
     \midrule
      Long TTI & 1\,ms   \\
     Total number of long TTI & $10^{5}$    \\
    Number of minislots (S) & 7    \\
    Number of subcarriers per RB  & 12    \\
    Number of OFDM symbols in long TTI ($N_{\text{sym}}$) & 14     \\
     Number of OFDM symbols in short TTI ($n_{\text{sym}}$) & 2     \\
    Subcarrier spacing  & 15 kHz    \\
    Bandwidth of each RB ($f_{b}$)  & 180\,kHz    \\
    Number of resource block (B)  & 4    \\
    Total system bandwidth (BW)  & 720 kHz    \\
     MC packet size ($\zeta$)  & 32 Byte    \\
    MC activation probability ($p_{m}$)  & 0.8    \\
    Power of MC user ($P_{m}$) & 30\,dBm
    \\
      Power of eMBB user ($P_{e}$) & 30\,dBm \\
      Tolerated loss in  average AoI  ($\beta$) & 0.1 minislot\\
      Threshold on PAoI  ($A_{\text{th}}$) & 4 minislot\\
   Tolerated loss in PAoI violation probability  ($\hat{\beta}$) &  0.01 \\
      SNR gap between eMBB and MC user ($\Gamma_{d}$) & $ \left [ -43, 43 \right ]$\,dB \\
      \bottomrule 
    \end{tabular}}
    \label{table.1}
\end{table}

\subsection{Performance Metrics of the MC User} 

\begin{figure*}[!t]
\centering{\includegraphics[width=1\linewidth]{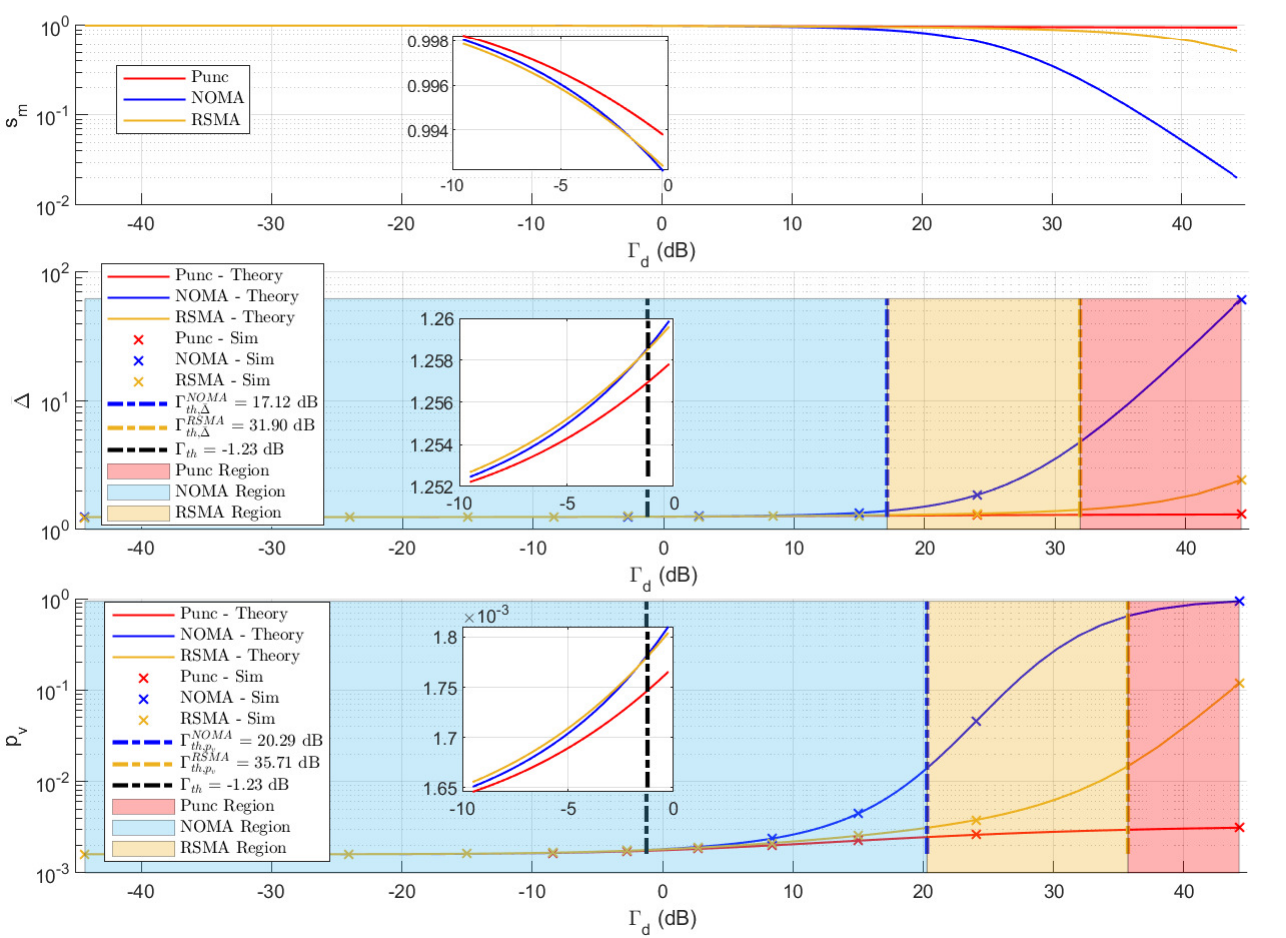}}
\caption{Probability of successful transmission ($s_m$), average AoI ($\bar{\Delta}$), and PAoI violation probability ($p_v$) for the MC user, as functions of the SNR gap ($\Gamma_d$).
}
\label{fig:6:sm-AAoI-PAoI-Sim-Theory}
\end{figure*}

Fig.~\ref{fig:6:sm-AAoI-PAoI-Sim-Theory} presents the probability of successful transmission ($s_m$), average AoI ($\overline{\Delta}$) and the PAoI violation probability ($p_\text{v}$) for the MC user when puncturing, NOMA, and RSMA are used.

Inspecting the behavior of $s_m$ in Fig.~\ref{fig:6:sm-AAoI-PAoI-Sim-Theory} (top) reveals that when the SNR gap (in dB) between the MC user and the eMBB user is up to some moderate to high values 
all three access schemes 
yield $s_m$ closer to the maximum, i.e., 1.
A notable difference among the schemes can be observed for $\Gamma_d$ larger than 15\,dB.
In terms of (only) $s_m$, the puncturing represents the optimal choice, as it shields the MC user from any interference emanating from the eMBB user.
NOMA performs second best up to the value of $\Gamma_{d}=\Gamma_{\text{th}} = - 1.23$\,dB, as shown in the zoomed-in area within the figure.
Specifically, before this point, there is a higher probability of successful decoding of the complete MC transmission in the NOMA, than of the gradual decoding of virtual users in the RSMA scheme, which is due to the robust channel gain of the MC user.
After this value of the SNR gap, a careful selection of the rate and power splitting factors $\lambda$ and $\omega$, respectively, among the virtual users according to the optimization problem presented in~\eqref{eq.OP.RSMA.a}, results in an overall higher successful decoding probability of the MC user. 

The middle part of Fig.\ref{fig:6:sm-AAoI-PAoI-Sim-Theory} depicts the average AoI ($\overline{\Delta}$) as a function of the SNR gap ($\Gamma_{d}$).
As anticipated, the average AoI behavior aligns with what could be expected from~\eqref{eq.4.average.AoI}: that is, higher values of $s_m$ correspond to lower average AoI, while lower $s_m$ values result in higher average AoI.
Consequentially, when $\Gamma_{d}$ falls below the threshold value of $\Gamma_{\text{th}} = -1.23$\,dB, it holds that $\overline{\Delta}_{\text{RSMA}} > \overline{\Delta}_{\text{NOMA}} > \overline{\Delta}_{\text{punc}}$.
For $\Gamma_{d} \geq \Gamma_{th}$, RSMA outperforms NOMA in terms of $s_m$ and, consequentially, $\bar{\Delta}_{\text{RSMA}} < \bar{\Delta}_{\text{NOMA}}$.
However, despite the RSMA's superior performance, NOMA still satisfies the inequality~\eqref{eq.OP.b.average.AoI} until $\Gamma_{\text{th},\bar{\Delta}}^{\text{NOMA}} = 17.12$\,dB.
Therefore, for $\Gamma_{d} < \Gamma_{\text{th},\bar{\Delta}}^{\text{NOMA}}$, NOMA remains the strategy that fulfills the criterion~\eqref{eq.OP.a.average.AoI} of maximizing the rate of the eMBB user.
Beyond this threshold, NOMA can no longer satisfy the condition~\eqref{eq.OP.b.average.AoI}, whereas RSMA can.
Thus, for $\Gamma_{\text{th},\bar{\Delta}}^{\text{NOMA}} \leq \Gamma_{d} < \Gamma_{\text{th},\bar{\Delta}}^{\text{RSMA}}$, RSMA emerges as the superior choice, where $\Gamma_{\text{th}, \bar{\Delta}}^{\text{RSMA}} = 31.90$\,dB.
Finally, if $\Gamma_{d} \geq \Gamma_{\text{th}, \bar{\Delta}}^{\text{RSMA}}$, neither NOMA nor RSMA can satisfy the condition in equation~\eqref{eq.OP.b.average.AoI}.
Hence, for $\Gamma_{d} \geq \Gamma_{\text{th}, \bar{\Delta}}^{\text{RSMA}}$, puncturing stands out as the optimal multiple access strategy, albeit at the expense of a significant decrease in the eMBB user's rate, as discussed later.

Fig.~\ref{fig:6:sm-AAoI-PAoI-Sim-Theory} (bottom) provides a comparative analysis of the PAoI violation probability ($p_v$) among the three schemes. 
Again, a reduction in the success probability $s_{m}$ leads to an increase in PAoI violation probability (c.f.,~\eqref{eq.6.PAoI}).
The findings reveal that at low SNR gaps (similar to the case for the average AoI) the PAoI violation probability follows the order $\overline{\Delta}_{\text{RSMA}} > \overline{\Delta}_{\text{NOMA}} > \overline{\Delta}_{\text{punc}}$ until $\Gamma_{d} = \Gamma_{\text{th}}$.
Consequently, NOMA emerges as the optimal multiple access scheme for $\Gamma_{d} < \Gamma_{\text{th}}$. After surpassing $\Gamma_{\text{th}}$, NOMA's performance in terms of PAoI violation probability $p_v$ gets worse compared to RSMA, yet not to an extent that \eqref{eq.OP.b.PAoI} cannot be satisfied. Thus, NOMA remains the preferred option up to the threshold $\Gamma_{\text{th},p_{v}}^{\text{NOMA}} = 20.29$\,dB.
Beyond $\Gamma_{\text{th},p_{v}}^{\text{NOMA}}$, NOMA fails to meet the constraint in~\eqref{eq.OP.b.PAoI}.
Consequently, RSMA becomes the optimal choice for the range $\Gamma_{\text{th},p_{v}}^{\text{NOMA}} \leq \Gamma_{d} < \Gamma_{\text{th},p_{v}}^{\text{RSMA}}$, where $\Gamma_{\text{th},p_{v}}^{\text{RSMA}} = 31.90$\,dB.
Lastly, if $\Gamma_{\text{th},p_{v}}^{\text{RSMA}} \geq \Gamma_{d}$, the puncturing is the preferred option, as neither of the other two schemes can satisfy \eqref{eq.OP.b.PAoI}.
The figure also shows that the SNR gap thresholds for NOMA and RSMA related to the average AoI and the PAoI violation probability have different optimal values. To ensure the satisfaction of both constraints in \eqref{eq.OP.b.average.AoI} and \eqref{eq.OP.b.PAoI},  the overall optimal thresholds can be determined according to~\eqref{threshold.optimal}.

{Fig.~\ref{fig:Optimal:omegaandlambda} illustrates the optimal power splitting factor for the MC user in RSMA ($\omega^{*}$), aiming to fulfill the optimization criteria in~\eqref{eq.OP.RSMA.a}.
Note that, as indicated by~\eqref{su.MC.last}, the success probability of RSMA remains unaffected by the activation probability of MC ($p_m$).
Consequently, the optimal values of power ($\omega^{*}$) and rate splitting factor ($\lambda^{*}$) remain independent of $p_m$.
When the SNR gap ($\Gamma_{d}$) is low, allocating higher power to virtual MC user 1 ($x_{m,1}$) is not needed for its perfect decoding; in this case, the high channel gain of the MC is sufficient to overcome interference from the eMBB user.
However, as the SNR gap between eMBB and MC increases, allocating more power to $x_{m,1}$ becomes essential to prevent interference from the eMBB user and ensure its successful decoding.

The evaluations also reveal that the optimal rate splitting factor ($\lambda^{*}$) maintains a consistent value across all SNR gaps, equivalent to the minimum permissible value of $\lambda$ in~\eqref{eq.OP.RSMA.c} which is 0.01. 
This is due to the fact that it is crucial to guarantee the successful decoding of ($x_{m,1}$) to achieve successful decoding of virtual MC user 2 ($x_{m,2}$).
Thus, a lower threshold on the MC rate is set for the first step, while a higher threshold is established for the third step of decoding (i.e., decoding $x_{m,2}$) as you can see in ~\eqref{eq.15.pout.MC.RSMA}. 

\subsection{Choosing the Optimal Transmission Strategy}

Fig.~\ref{fig:7:Re-Sim-Theory} illustrates the achievable data rate for the eMBB user in puncturing, NOMA, and RSMA based on \eqref{eq.8.Re.punc}, \eqref{eq.13.Re.NOMA}, and \eqref{eq.Re.RSMA} from Sec.~\ref{Analysis}.
Notably, the achievable data rate for the eMBB user with puncturing is the lowest, due to preemption of its transmission when the MC user is active.

In \eqref{eq.13.Re.NOMA} that corresponds to the NOMA scheme, when the MC user is active, there are two possible scenarios: either there is no interference from the MC user to the eMBB user when the MC user is successfully decoded, or there is a complete interference from the MC user to the eMBB user when the MC user is not decoded.
In the RSMA scheme, \eqref{eq.Re.RSMA} also shows two alternatives: there can be partial interference from the MC user to the eMBB user when virtual user 1 ($x_{m,1}$) is successfully decoded, or complete interference from the MC to the eMBB user when $x_{m,1}$ is not decoded.
When $\Gamma_{d} < -1.23$\,dB, the successful probability of decoding of the MC user is nearly 1, as evident from Fig.~\ref{fig:6:sm-AAoI-PAoI-Sim-Theory} (top).
This leads to the elimination of all interference from the received signal in the NOMA scenario.
However, in the RSMA scheme, there is always a partial interference to the eMBB user caused by virtual user 2 ($x_{m,2}$).
Consequently, the achievable data rates for eMBB in RSMA and puncturing show similar behavior, as more power is allocated to $x_{m,2}$ compared to $x_{m,1}$ (as can be seen in Fig.~\ref{fig:Optimal:omegaandlambda}).
When $\Gamma_{d} \geq -1.23$\,dB, interference from the eMBB user starts to affect the decoding process of the MC user.
Thus, the achievable data rates for the eMBB user in NOMA and RSMA start approaching each other.

By calculating $\Gamma_{\text{th}}^{*,\text{NOMA}}$ and $\Gamma_{\text{th}}^{*,\text{RSMA}}$ from \eqref{threshold.optimal}, operating ranges of $\Gamma_d$ can be derived corresponding to the optimal choice among the puncturing, NOMA, or RSMA.
Such a decision ensures the target average AoI and PAoI violation probability for the MC user while maximizing the data rate for the eMBB user.
The performance of the adaptive scheme derived using this approach is depicted by the black circles in Fig.~\ref{fig:7:Re-Sim-Theory}.




\begin{figure}[!t]
\centering{\includegraphics[trim={0.20cm 0.25cm 0.2cm 0.18cm},clip, width=1\linewidth]{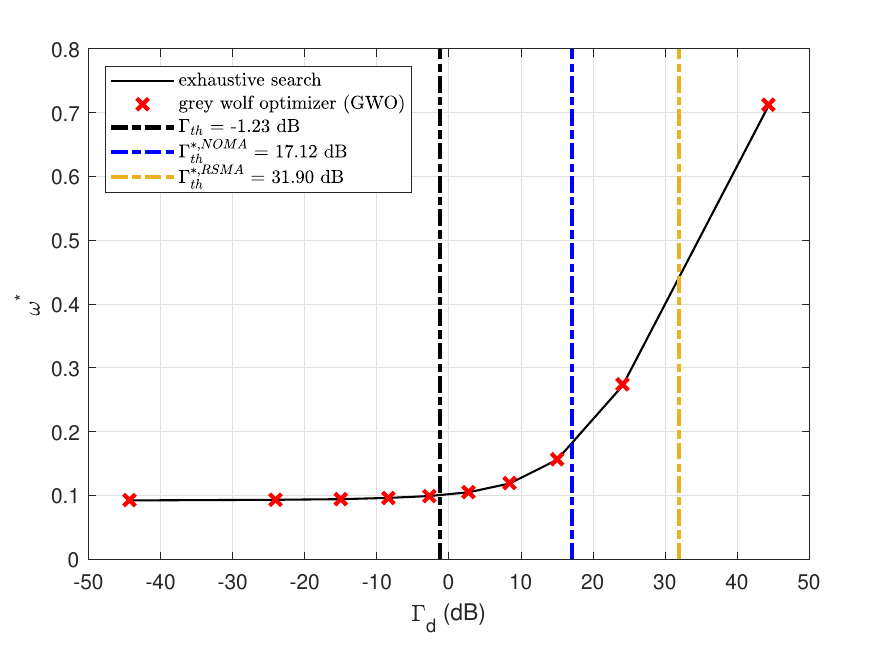}}
\caption{Optimal power splitting factor ($\omega^{*}$) for virtual MC user 1 in RSMA to maximize the overall success probability of MC ($s_{m}$) with exhaustive search and grey wolf optimizer (GWO) method w.r.t. SNR gap ($\Gamma_d$), see~\eqref{eq.OP.RSMA.a}.
}
\label{fig:Optimal:omegaandlambda}
\end{figure}
\begin{figure}[!t]
\centering{\includegraphics[trim={0.32cm 0.35cm 1.3cm 0.18cm},clip, width=1\linewidth]{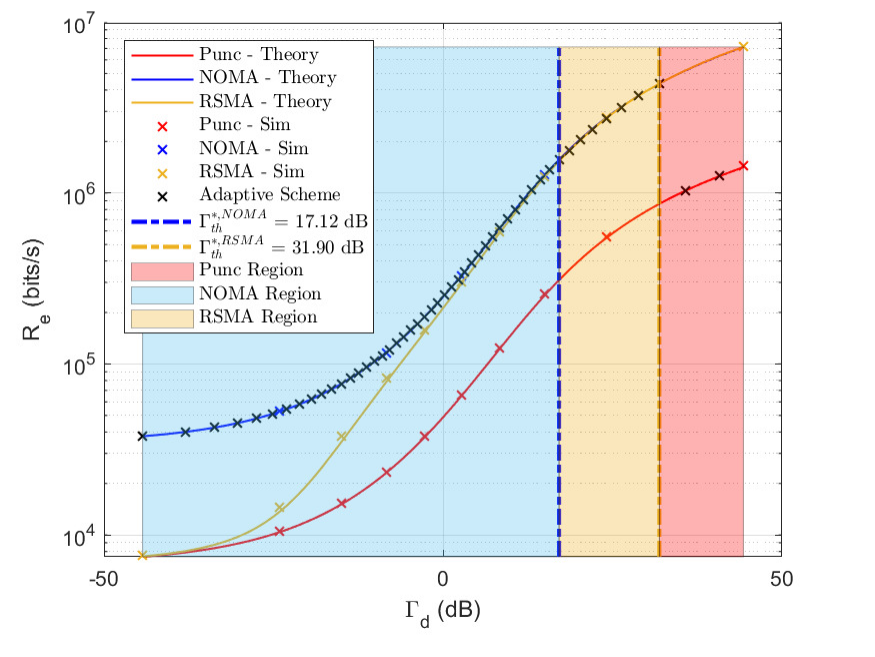}}
\caption{Achievable rate for eMBB user ($R_{e}$) as a function of the SNR gap ($\Gamma_d$).
}
\label{fig:7:Re-Sim-Theory}
\end{figure}
\subsection{Dependence of $\Gamma_{\text{th}}^{*,\text{NOMA}}$ and $\Gamma_{\text{th}}^{*,\text{RSMA}}$ on the Activation Probability of the MC User}
\begin{figure}[!t]
\centering{\includegraphics[trim={0.32cm 0.35cm 1.3cm 0.18cm}, width=1 \linewidth]{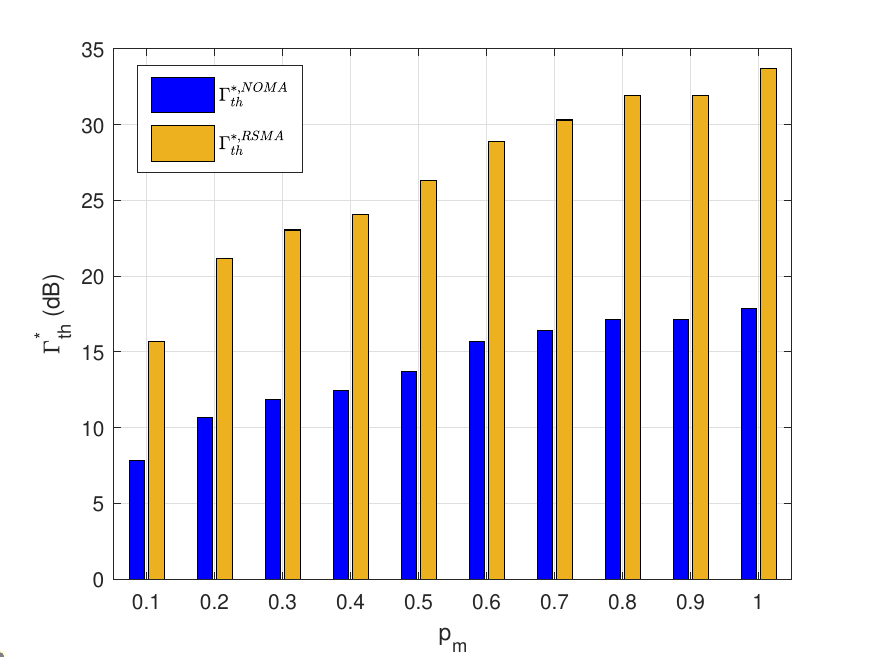}}
\vspace{-8pt}
\caption{$\Gamma_{\text{th}}^{*,\text{NOMA}}$ and $\Gamma_{\text{th}}^{*,\text{RSMA}}$ as a function of the activation probability of the MC user $p_m$; $\beta = 0.1$~minislot and $\hat{\beta} = 0.01$.}
\label{fig:5:Threshold:Activation:Probability}
\end{figure}

Fig.~\ref{fig:5:Threshold:Activation:Probability} shows the behavior of $\Gamma_{\text{th}}^{*,\text{NOMA}}$ and $\Gamma_{\text{th}}^{*,\text{RSMA}}$ as a function of different activation probabilities ($p_{m}$) of the MC user.
When $p_m$ is low, the MC user transmits sporadically and the individual packet losses significantly impact the average AoI ($\bar{\Delta}$) and the PAoI violation probability ($p_v$).
Consequently, a high success probability ($s_{m}$) is required, dictating a low tolerable SNR gap between the eMBB and the MC user.

As $p_m$ increases, $s_m$ does not need to be as high to keep $\bar{\Delta}$ and $p_v$ close to the best possible values, and the SNR gap $\Gamma_{d}$ that can be tolerated grows.
The results indicate that $\Gamma_{\text{th}}^{*,\text{RSMA}}$ is roughly two times higher than $\Gamma_{\text{th}}^{*,\text{NOMA}}$ for the selected configuration parameters.
This can be intuitively explained by considering that in the RSMA framework, there exists a degree of freedom to adjust the rate splitting factor ($\lambda$) and the power splitting factor ($\omega$) for the virtual MC users.
These two factors are optimized to fulfill the constraints outlined in~\eqref{eq.OP.b.average.AoI} and \eqref{eq.OP.b.PAoI}, in effect contributing to the overall increase in $s_m$.
On the other hand, in NOMA, the MC user must always be decoded before the eMBB user.
This restriction imposes a lower threshold on the constraints presented in equations~\eqref{eq.OP.b.average.AoI} and \eqref{eq.OP.b.PAoI} in comparison to the RSMA scenario.



\section{Conclusion}
\label{Conclusion}
In this paper, we examined RSMA, NOMA, and puncturing access methods in an uplink transmission framework accommodating the coexistence of MC and eMBB users. Our analysis focused on the average AoI and PAoI violation probability for MC user, as well as the achievable data rate for eMBB user. The close form of the average AoI and PAoI violation probability is derived by using the DTMC model and determining the
PAoI distribution, respectively. Our motivation was to identify the conditions under which NOMA/RSMA can be applied with minimal impact on MC performance, considering the SNR gap between the eMBB and the MC users. In addition, the optimal power and rate splitting factors for RSMA are derived to maximize the success probability of the MC user. Based on the insights from analytical/simulation results, we showed that an adaptive access strategy that dynamically switches between RSMA, NOMA, and puncturing methods based on the channel gain difference between eMBB and MC users can be exploited to achieve the requisite average AoI and PAoI violation probability performance while maximizing the data rate for the eMBB user.

The proposed access strategy, developed based on the practical performance metrics for conflicting QoS requirements of the MC and eMBB users, provides actionable insights into the trade-offs associated with different access schemes with respect to their critical parameters. These insights are useful for the radio access design in many industrial application domains presenting mixed-traffic types, such as (but not limited to) Industrial IoT (IIoT), smart grids, and vehicle-to-everything (V2X).  To strengthen the applicability of our proposed solution further in practical scenarios, in future work, we aim to study multiple eMBB and MC users' coexistence scenarios, exploring sophisticated scheduling algorithms that take into account the SNR gap, and optimizing the selection of the most suitable multiple access scheme (RSMA/NOMA/puncturing) to enhance overall system performance.

\bibliographystyle{IEEEtran}
\bibliography{main.bib}

\begin{IEEEbiography}[{\includegraphics[width=1in, height=1.25in,clip, keepaspectratio]{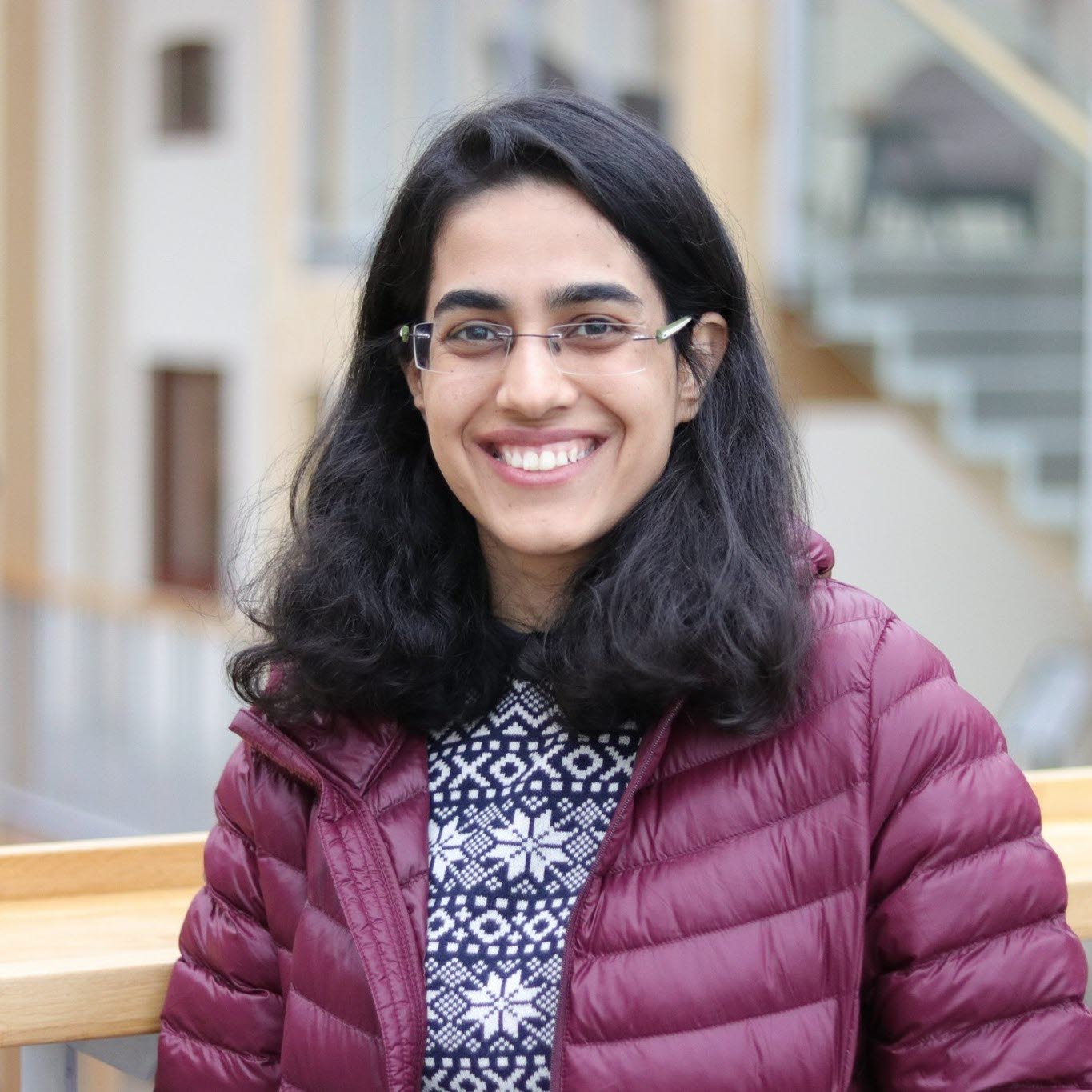}}] {Farnaz~Khodakhah} (Student Member, IEEE) received the B.E. degree in electrical engineering, communications from the Shiraz University of Technology, Shiraz, Iran, in 2015, and the M.Sc. degrees in communications engineering, system from Shiraz University, Shiraz, Iran, in 2019. She is currently a Ph.D. at Mid Sweden University and her main research focuses on wireless
medium access for reliable IoT connectivity.

\end{IEEEbiography}

\begin{IEEEbiography}[{\includegraphics[width=1in, height=1.25in,clip, keepaspectratio]{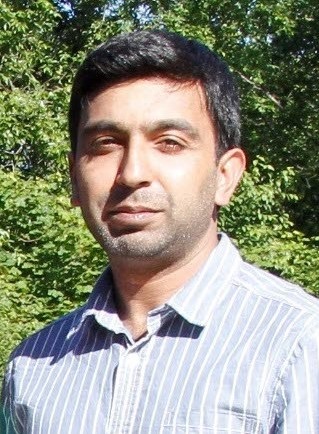}}] {Aamir~Mahmood} (Senior Member, IEEE) received the B.E. degree in electrical engineering from the National University of Sciences and Technology, Islamabad, Pakistan, in 2002, and the M.Sc. and D.Sc. degrees in communications engineering from the Aalto University School of Electrical Engineering, Espoo, Finland, in 2008 and 2014, respectively. He worked as a Research Intern with Nokia Research Center, Helsinki, Finland, in 2014, as a Visiting Researcher with Aalto University from 2015 to 2016, and as a Postdoc with Mid Sweden University, Sundsvall, Sweden, from 2016 to 2018, where he has been an Associate Professor with the Department of Computer and Electrical Engineering, since 2023. His research interests include Industrial IoT, 5G-TSN integration, AI/ML for radio network optimization and management, RF interference and coexistence management, network time synchronization, and wireless positioning.
\end{IEEEbiography}

\begin{IEEEbiography}[{\includegraphics[width=1in, height=1.25in,clip, keepaspectratio]{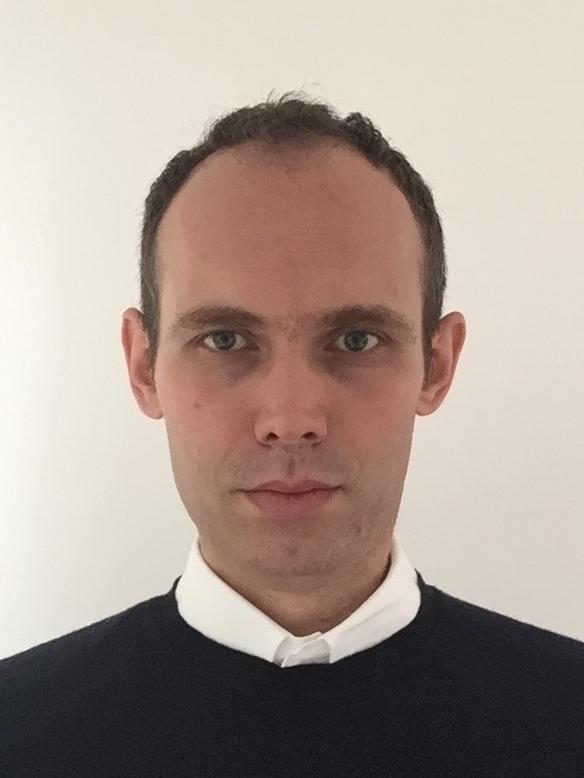}}] {\v{C}edomir~Stefanovi\'{c}} (Senior Member, IEEE)
received the Diploma Ing., Mr.-Ing., and Ph.D.
degrees from the University of Novi Sad, Serbia.
He is currently a Professor with the Department
of Electronic Systems, Aalborg University, where
he leads Edge Computing and Networking Group.
He is a Principal Researcher on a number of
European projects related to IoT, 5G, and mission-critical communications. He has co-authored more
than 130 peer-reviewed publications. His research
interests include communication theory and wireless communications. 
\end{IEEEbiography}

\begin{IEEEbiography}[{\includegraphics[width=1in, height=1.25in,clip, keepaspectratio]{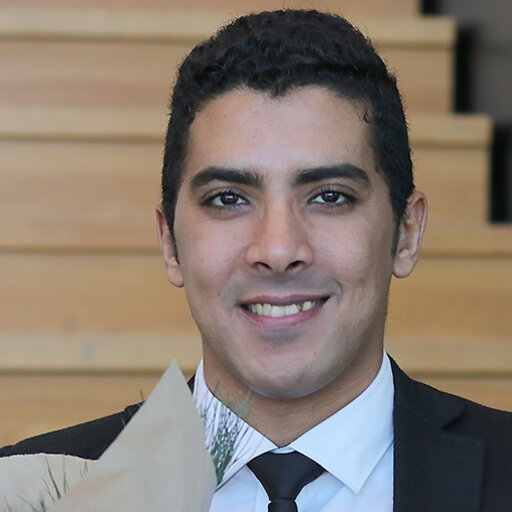}}] {Hossam Farag} (Member, IEEE) received M.Sc. degree in Electrical Engineering from Aswan University (Egypt) in
2015 and PhD degree in Computer Engineering
from Mid Sweden University (Sweden) in 2020.
From 2020 to 2022, he was a postdoctoral fellow
with the Department of Electronic Systems at
Aalborg University (Denmark). Currently, he is
an Assistant Professor at the Department of
Electronic Systems at Aalborg University (Denmark). His current research focuses on random
access protocols, wireless sensor networks, cellular networks, machine learning for wireless networks, reliable and realtime communication for IoT networks.

\end{IEEEbiography}

\begin{IEEEbiography}[{\includegraphics[width=1in, height=1.25in,clip, keepaspectratio]{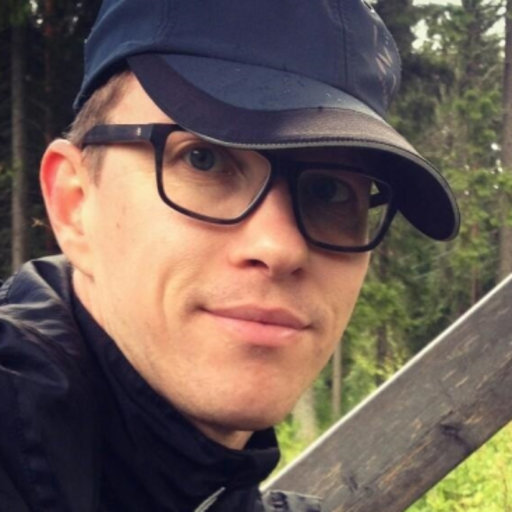}}] {Patrik~{\"O}sterberg} (Senior Member, IEEE) received the
M.Sc. degree in electrical engineering from Mid
Sweden University, Sundsvall, Sweden, in 2000,
the degree of licentiate of technology in teleinformatics from the Royal Institute of Technology,
Stockholm, Sweden, in 2005, and the Ph.D. degree in computer and system science from Mid
Sweden University, in 2008.
In 2007, he worked as a Development Engineer with Acreo AB, Hudiksvall, Sweden. From
2008 to 2010, he was employed as a Researcher with Interactive TV Arena KB in G{\"a}vle, Sweden. He has been an Associate Professor at Mid Sweden University since 2022 and the Head of the Department of Computer and Electrical Engineering since 2013.

\end{IEEEbiography}

\begin{IEEEbiography}[{\includegraphics[width=1in, height=1.25in,clip, keepaspectratio]{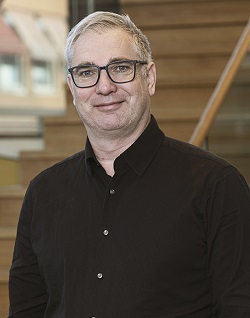}}]{Mikael Gidlund} (Senior Member, IEEE) received the Licentiate of Engineering degree in radio communication systems from the KTH Royal Institute of Technology, Stockholm, Sweden, in 2004, and the Ph.D. degree in electrical engineering from Mid Sweden University, Sundsvall, Sweden, in 2005. From 2008 to 2015, he was a Senior Principal Scientist and Global Research Area Coordinator of Wireless Technologies with ABB Corporate Research, V\(\ddot{\text{a}}\)ster\(\mathring{\text{a}}\)s, Sweden. From 2007 to 2008, he was a Project Manager and a Senior Specialist with Nera Networks AS, Bergen, Norway. From 2006 to 2007, he was a Research Engineer and a Project Manager with Acreo AB, Hudiksvall, Sweden. Since 2015, he has been a Professor of Computer Engineering at Mid Sweden University. He holds more than 20 patents (granted and pending) in the area of wireless communication. His current research interests include wireless communication and networks, wireless sensor networks, access protocols, and security. Dr. Gidlund is an Associate Editor of the {\it IEEE Transactions on Industrial Informatics.}  \end{IEEEbiography}
\end{document}